\documentclass[aps,prd,10pt,showpacs]{revtex4-1}
\usepackage[english]{babel}
\usepackage{amsmath, amsfonts, amssymb, bm, amsthm} 
\usepackage{graphicx}
\usepackage{multirow}

\newcommand{\Diff}[2]{\frac{\text{d} #1}{\text{d} #2}}

\renewcommand{\vec}{\mathbf}

\newcommand{\fract}[2]{{#1}\slash{#2}}

\newcommand{\gag}{g_{a\gamma}}
\newcommand{\ma}{m_{a}}

\newcommand{\apjs}{Astrophys. J. Suppl. Ser.}
\newcommand{\aap}{Astron. Astrophys.}
\newcommand{\apjl}{Astrophys. J., Lett.}

\newcommand{\mnras}{Mon. Not. R. Astron. Soc.}
\newcommand{\araa}{Ann. Rev. Astron. Astrophys.}

\newcommand{\jcap}{JCAP}
\newcommand{\aapr}{Astron. Astrophys. Review}

\newcommand{\hess}{H.E.S.S.~}

\newcommand{\scenIGMF}{\emph{optimistic IGMF}}
\newcommand{\scenICM}{\emph{optimistic ICM}}
\newcommand{\scenfid}{\emph{fiducial}}
\newcommand{\scenmax}{\emph{general source}}
\newcommand{\LL}{0.01}
\newcommand{\logLL}{2}

\newcommand{\gray}{$\gamma$-ray}
\newcommand{\BIGMF}{B_\mathrm{IGMF}}
\newcommand{\CIGMF}{\lambda_\mathrm{IGMF}^c}
\newcommand{\BICMF}{B_\mathrm{ICMF}}
\newcommand{\CICMF}{\lambda_\mathrm{ICMF}^c}

\newcommand{\medphotsurv}{\tilde{P}_{\gamma\gamma}}
\newcommand{\KD}{KD model}
\newcommand{\FRV}{FRV model}
\newcommand{\pa}{photon-ALP}
\newcommand{\Pa}{Photon-ALP}

\begin{document}

\title{First lower limits on the photon-axion-like particle coupling from very high energy \gray~observations} 

\author{Manuel Meyer}
\email[]{manuel.meyer@physik.uni-hamburg.de}
\author{Dieter Horns}
\author{Martin Raue}
\affiliation{Institut f\"ur Experimentalphysik, Universit\"at Hamburg, Luruper Chaussee 149, 22761 Hamburg, Germany}

\date{\today}

\begin{abstract}

The intrinsic flux of very high-energy (VHE, energy $\gtrsim 100$\,GeV) \gray s from extragalactic sources is attenuated due to pair production in the interaction with photons of
 the extragalactic background light (EBL).
Depending on the distance of the source, the Universe should be opaque to VHE photons above a certain energy.
However, indications exist that the Universe is more transparent than previously thought. 
A recent statistical analysis of a large sample of VHE spectra shows that the correction for absorption with current EBL models is too strong for the 
data points with the highest attenuation.
An explanation might be the oscillation of VHE photons into hypothetical axionlike particles (ALPs) in ambient magnetic fields.
This mechanism would decrease the opacity, as ALPs propagate unimpeded over cosmological distances.

Here, a large sample of VHE \gray~spectra obtained with imaging air Cherenkov telescopes is used to set, for the first time, 
lower limits on the \pa~coupling constant $\gag$ over a large range of ALP masses.
The conversion in different magnetic field configurations, including intra-cluster and intergalactic magnetic fields together with the magnetic field of the Milky Way, is investigated 
taking into account the energy dependence of the oscillations.
For optimistic scenarios of the intervening magnetic fields, a lower limit on $\gag$ of the order of $10^{-12}\mathrm{GeV}^{-1}$ is obtained,
whereas more conservative model assumptions result in $\gag \gtrsim 2\times10^{-11}\mathrm{GeV}^{-1}$.
The latter value is within reach of future dedicated ALP searches.
\end{abstract}


\maketitle 

\section{Introduction}
\label{sec:intro}

Very high-energy (VHE, energy $\gtrsim 100$\,GeV) \gray s from extragalactic sources interact with photons of the extragalactic background light (EBL) and produce $e^+e^-$ pairs \cite{nikishov1962,jelley1966,gould1966}.
As a result, the intrinsic \gray~flux is exponentially suppressed with the optical depth, $\tau(z,E)$, that increases with both the redshift $z$ of the source and energy $E$.
Above a certain energy, the Universe should thus become opaque to \gray s from sufficiently distant sources.
The optical depth also depends on the photon number density of the EBL whose exact level remains unknown,
 as direct observations are difficult due to contamination by foreground emission \cite{hauser1998}.
The EBL ranges from ultraviolet (UV) / optical to far-infrared wavelengths and originates from the starlight integrated over all epochs and the starlight absorbed and reemitted by dust in galaxies
(see e.g. Refs. \cite{hauser2001, dwek2012} for a review).
Nevertheless, firm lower limits in the optical and infrared exist which can be derived from galaxy number counts \cite{madau2000,fazio2004},
and recent EBL models show a good agreement in the overall shape of the spectrum (e.g. Refs. \cite{franceschini2008, kneiske2010, dominguez2011}).
In addition, VHE-\gray~spectra from extragalactic sources can be used to place upper limits on the EBL density 
if certain assumptions about the intrinsic spectra are made (see e.g. Refs. \cite{orr2011,meyer2012} for recent upper limits).

However, indications have been found that the 
Universe is more transparent to VHE \gray s than the prediction of EBL models suggests
(e.g. Refs. \cite{1es0229hess2007,3c279magic2008,deangelis2007,mirizzi2007}).
The authors of Ref. \cite{horns2012} introduced a nonparametric test to quantify the accordance between EBL model predictions and VHE observations. 
They tested the EBL model of Ref. \cite{kneiske2010} (henceforth the \KD)
which is designed to produce a minimal attenuation as it closely follows the galaxy number counts in the infrared.
The authors used a statistical sample of the spectra of active galactic nuclei (AGNs) measured with imaging air Cherenkov telescopes (IACTs) 
and fitted analytical functions to the spectral points \cite{deangelis2009}.
Subsequently, the authors tested if the residuals of data points that correspond to an optical depth $\tau \geqslant 2$  follow the expectation to scatter around a zero mean.
They found, though, that the mean is larger than zero at a $\sim 4\,\sigma$ significance level, indicating an overcorrection of the data with the \KD.
A possible solution to this observed pair production anomaly might be the conversion of photons into axionlike particles (ALPs) \cite{deangelis2007,mirizzi2007}.
ALPs are pseudo Nambu-Goldstone bosons that are created if additional global symmetries to the standard model are spontaneously broken (see e.g. Ref. \cite{jaeckel2010} for a review).
Such fields are a common prediction in compactified string theories \cite[][]{cicoli2012}.
Their phenomenology is closely related to that of axions which solve the strong CP problem in QCD \cite{pq1977,weinberg1978,wilczek1978}.
Most importantly in the present context, ALPs share the same coupling to photons as axions, 
characterized by the Lagrangian
\begin{equation}
 \mathcal{L}_{a\gamma} = \frac 1 4 \gag \,\tilde{F}_{\mu\nu}F^{\mu\nu} a = \gag \,\vec{E}\cdot\vec{B}\, a,
 \label{Eq:gag}
\end{equation}
where $F^{\mu\nu}$ is the electromagnetic field tensor (with electric and magnetic fields $\vec{E}$ and $\vec{B}$, respectively),
 $\tilde{F}^{\mu\nu}$ is its dual, $a$ is the ALP field strength, and $\gag$ is the \pa~coupling strength which has the dimension $(\mathrm{Energy})^{-1}$.
In contrast to the axion, the ALP mass $m_a$ is unrelated to the coupling strength.

The effect of \pa~mixing on VHE \gray~spectra in different magnetic field settings has been extensively discussed in the literature. 
The case of a conversion in an intergalactic magnetic field (IGMF) was addressed by, e.g., Refs. \cite{deangelis2007,mirizzi2007,mirizzi2009,deangelis2011,wouters2012}. 
The authors of, e.g., Refs. \cite[][]{sanchezconde2009,tavecchio2012} included the magnetic fields in and around the source,
 and in Ref. \cite{simet2008} the backconversion of ALPs into photons in the galactic magnetic field (GMF) of the Milky Way was studied.
Recently, the \pa~mixing for sources located inside galaxy clusters and the reconversion in the GMF was also investigated \cite{horns2012ICM}.
Usually, previous studies used fixed values for the ALP mass and coupling close to current experimental bounds in order to maximize the effect on the \gray~spectra.

In this article, the goal is to determine the preferred region in the $(\ma,\gag)$ parameter space
that minimizes the tension between data and model predictions found in Ref. \cite{horns2012}. 
This allows, for the first time, to place a lower limit on the \pa~coupling to explain the observed transparency of the Universe for VHE \gray s.
Four different scenarios for the intervening magnetic field will be considered, including mixing in the IGMF, the intracluster magnetic field (ICMF), and the GMF of the Milky Way.
In two cases, the parameters of the IGMF and ICMF will be chosen as optimistically as possible in order to derive lower limits on $\gag$.
Additionally, a more conservative choice of $B$-field model parameters will be investigated.
Furthermore, two different EBL model realizations will be studied. 

The paper is organized as follows:
In Sec. \ref{sec:photonALP}, the theory of \pa~conversions is briefly reviewed.
In Sec. \ref{sec:Bfield}, the different magnetic field configurations and the scenarios that are analyzed are summarized.
Subsequently, the method of constraining the ALP parameter space is laid out in Sec. \ref{sec:method} before presenting the actual results in Sec. \ref{sec:results}
and concluding in Sec. \ref{sec:conclusion}.

\section{\Pa~conversion in magnetic fields}
\label{sec:photonALP}
The \pa~interaction is described by the Lagrangian
\begin{equation}
 \mathcal{L} =  \mathcal{L}_{a\gamma} + \mathcal{L}_\mathrm{EH} + \mathcal{L}_a
\end{equation}
where $\mathcal{L}_a$ is given in Eq. (\ref{Eq:gag}), $\mathcal{L}_\mathrm{EH}$ is the effective Euler-Heisenberg Lagrangian accounting for one-loop corrections in the photon propagator (see, e.g., Ref. \cite[][]{raffelt1988})
and $\mathcal{L}_a$ includes the kinetic and mass term of the ALP,
\begin{equation}
 \mathcal{L}_{a} = \frac 1 2 \partial_\mu a \partial^\mu a - \frac 1 2 \ma^2 a^2.
 \label{Eq:kin}
\end{equation}
For a monochromatic photon / ALP beam of energy $E$ propagating along the $x_3$ axis in a cold plasma with a homogeneous magnetic field,  it can be shown that $\mathcal{L}$
leads to the following Schr\"odinger-like equation of motion \cite{raffelt1988}:
\begin{equation}
 \left( i\Diff{}{x_3} + E + \mathcal{M}_0 \right) \begin{pmatrix}
                                                 A_1(x_3) \\ A_2(x_3) \\ a(x_3)
                                                \end{pmatrix}
= 0,
\label{eq:eom}
\end{equation}
where $A_1(x_3)$ and $A_2(x_3)$ describe the linear photon polarization amplitudes along $x_1$ and $x_3$, respectively, 
and $a(x_3)$ denotes the ALP field strength. Let $\vec{B}_\perp$ denote the magnetic field transverse to the beam propagation direction.
If one chooses $\vec{B}_T$ to lie only along the $x_2$ direction, the mixing matrix $\mathcal{M}_0$ can be written as
\begin{equation}
 \mathcal{M}_0 = \begin{pmatrix}
                  \Delta_\perp	&		0	&			0\\
		  0		&	\Delta_{||}	&	\Delta_{a\gamma} \\
		  0		&	\Delta_{a\gamma}&	\Delta_{a}	 \\
                 \end{pmatrix},
\end{equation}
where a mixture of the photon polarization states due to Faraday rotation can be safely neglected for the energies considered here.
The matrix elements $\Delta_{||} = \Delta_\mathrm{pl} + 7/2 \Delta_\mathrm{QED} $ and $\Delta_\perp =  \Delta_\mathrm{pl} + 2\Delta_\mathrm{QED}$ account for medium effects on the photon propagation, 
where $\Delta_\mathrm{pl} = -\omega_\mathrm{pl} / (2E)$ with the plasma frequency of the medium, $\omega_\mathrm{pl}$. 
The plasma frequency is connected to the ambient thermal electron density $n_\mathrm{el}$ through $\omega_\mathrm{pl} = 3.69\times10^{-11}\sqrt{n_\mathrm{el}/\mathrm{cm}^{-3}}$\,eV.
The QED vacuum birefringence effect is included in $\Delta_\mathrm{QED} \propto B_\perp^2$.
Furthermore, $\Delta_{a} = -\ma^2/(2E)$ and the \pa~mixing is induced by the off-diagonal element $\Delta_{a\gamma} = 1/2 \gag B_\perp$ 
(see, e.g., \cite[][]{horns2012ICM} for numerical values in suitable units of the matrix elements).
If photons are lost due to the interaction with the EBL, the elements $\Delta_{||,\perp}$ are modified to include a complex absorption term,
$\Delta_{||,\perp} \to \Delta_{||,\perp} + i / (2 \lambda^\mathrm{mfp}_\gamma)$, where $\lambda^\mathrm{mfp}_\gamma$ is the mean free path for photons undergoing pair production.
Equation (\ref{eq:eom}) is solved with the transfer function $\mathcal{T}(x_3,0;E)$ with the initial condition $\mathcal{T}(0,0;E) = 1$.
Neglecting the birefringence contribution for a moment, it can be shown that the \pa~oscillations become maximal and independent of the energy $E$ and ALP mass $\ma$
 for an energy above the critical energy 
\begin{equation}
E_\mathrm{crit} \equiv E \frac{|\Delta_a - \Delta_\mathrm{pl}|}{2\Delta_{a\gamma}} 
  \approx 2.5 \frac{|m_a^2 - \omega_\mathrm{pl}^2|}{1\,\mathrm{neV}} \left(\frac{\gag}{10^{-11}\mathrm{GeV}^{-1}}\right)^{-1}\left(\frac{B_\perp}{1\,\mu\mathrm{G}}\right)^{-1} \,\mathrm{GeV}
\label{eq:Ecrit}
\end{equation}
 which is the so-called strong mixing regime. 
However, as the goal of this paper is to constrain the $(\ma,\gag)$ parameter space, it is generally not the case that the mixing occurs in this regime.

So far, only a polarized photon beam has been considered. As of today, the polarization of VHE \gray s cannot be measured and one has to consider an unpolarized photon beam 
and reformulate the problem in terms of density matrices. The general polarization matrix is given by
\begin{equation}
 \rho(x_3) = 
\begin{pmatrix}
    A_1(x_3) \\ A_2(x_3) \\ a(x_3)
\end{pmatrix}
\otimes
\begin{pmatrix}
    A_1(x_3) & A_2(x_3) & a(x_3)
\end{pmatrix}^\ast
\end{equation}
and the equation of motion takes the form of a von Neumann-like equation,
\begin{equation}
 i\Diff{\rho}{x_3} = [\rho,\mathcal{M}_0],
\label{eq:vneu}
\end{equation}
which is solved by $\rho(x_3,E) = \mathcal{T}(x_3,0;E)\,\rho(0)\,\mathcal{T}^\dagger(x_3,0;E)$.
In the more general case in which $\vec{B}_\perp$ has an arbitrary orientation and forms an angle $\psi$ with the $x_2$ axis, the solution can be found via a similarity transformation
\begin{equation}
V(\psi) = \begin{pmatrix}
                  \cos\psi	&	-\sin\psi	&	0\\
		  \sin\psi	&	\cos\psi	&	0\\
		  0		&		0	&	1\\
                 \end{pmatrix},
\end{equation}
so that $\mathcal{M} = V(\psi)\mathcal{M}_0 V^\dagger(\psi)$ and the solution to the modified Eq. (\ref{eq:vneu}) is $\mathcal{T}(x_3,0;E;\psi) = V(\psi) \mathcal{T}(x_3,0;E) V^\dagger(\psi)$.
If, moreover, the beam path can be split up into $n$ domains with a constant and homogeneous magnetic field in each domain but a changing orientation (and strength) from one domain to the next,
the complete transfer matrix is simply given by the product over all domains,
\begin{equation}
 \mathcal{T}(x_{3, n},x_{3, 0};E;\psi_{n-1},\ldots,\psi_0) = \prod\limits_{k = 0}^{n-1} \mathcal{T}_k(x_{3, k + 1},x_{3, k};E;\psi_k),
\end{equation}
with one mixing matrix $\mathcal{M}_k$ for each domain. 
The transition probability of observing a photon / ALP beam in the state $\rho_\mathrm{final}$ after the crossing of $n$ magnetic domains reads
\begin{equation}
P_\mathrm{final} = \mathrm{Tr}(\rho_\mathrm{final} \mathcal{T}(x_{3, n},x_{3, 0};E;\psi_{n-1},\ldots,\psi_0) \rho(x_{3,0}) \mathcal{T}^\dagger(x_{3, n},x_{3, 0};E;\psi_{n-1},\ldots,\psi_0)).
\label{eq:transprob}
\end{equation}
Equipped with this formula, the photon transition probability $P_{\gamma\gamma}$ is defined as the sum of the transition probabilities
from an initially unpolarized pure photon state $\rho_\mathrm{unpol} = 1/2 \mathrm{diag}(1,1,0)$ to 
the final polarization states $\rho_{11} = \mathrm{diag}(1,0,0)$ and $\rho_{22} = \mathrm{diag}(0,1,0)$: 
\begin{equation}
 P_{\gamma\gamma} = P_{11} + P_{22} = \mathrm{Tr}((\rho_{11} + \rho_{22})\mathcal{T}\rho_\mathrm{unpol}\mathcal{T}^\dagger).
\label{eq:phot-surv}
\end{equation}

\section{Magnetic field configurations and scenarios}
\label{sec:Bfield}

When the photon / ALP beam propagates towards Earth, it crosses different regions of plasma and magnetic field configurations. 
The following environments are considered, ordered by increasing distance from Earth:
\begin{enumerate}
 \item The Galactic magnetic field of the Milky Way (GMF).
 \item The intergalactic magnetic field (IGMF).
 \item The magnetic field inside a galaxy cluster (intracluster magnetic field, ICMF) in the vicinity of the emitting source.
\end{enumerate}
The observational evidence and model assumptions for each region are discussed in the following subsections. The goal is to find the magnetic field configuration within current observational bounds 
which results in a maximal \pa~mixing. In this way, a lower limit on $\gag$ can be derived.

\subsection{Magnetic field of the Milky Way}
The regular component of the $B$ field of the Milky Way will be described with the analytical GMF model presented in \cite{jansson2012}.
The model consists of three components: a disk, a halo, and a so-called X component; and it predicts a field strength of the order $\mathcal{O}(\mu\mathrm{G})$.
The model parameters were determined with a $\chi^2$ minimization utilizing the data of the WMAP7 polarized synchrotron emission maps and Faraday rotation measurements of extragalactic sources.
Compared to previous models (e.g. Ref. \cite{pshirkov2011}), a relatively large field strength and extent is predicted for the halo and X component
which leads to a comparatively large conversion probability in certain regions in the sky.
The turbulent component of the GMF is neglected here, since the typical coherence length is of the order of $\mathcal{O}(10\,\mathrm{pc})$ which is far smaller than the oscillation length of \pa-conversions.
For each extragalactic VHE-\gray~source, the conversion probability is evaluated along the line of sight where it is assumed that the GMF is constant and homogeneous on a length scale of 100\,pc. 
It was checked that smaller values for the domain length do not alter the results.
Moreover, the density of the thermal electron plasma is calculated with the NE2001 code (in accordance with Ref. \cite{jansson2012}) which predicts densities of the order of $10^{-1}\,\mathrm{cm}^{-3}$ \cite{ne2001}.
For further details on the conversion in the GMF, see Ref. \cite{horns2012ICM}, where the same GMF model was utilized to compute the \pa~conversions.

\subsection{Intergalactic magnetic field}
In contrast to the GMF, little is known about the intergalactic magnetic field. 
From the observational side, only upper limits exist on the field strength, 
which constrain the IGMF at $z = 0$ to a few $10^{-9}\,$G \cite{kronberg1994}; and, e.g, the authors of Ref.
\cite{blasi1999} find $\BIGMF^0 \equiv \BIGMF(z = 0) \lesssim 6 \times 10^{-9}$\,G 
for a coherence length of $\CIGMF= 50$\,Mpc using Faraday rotation measurements of quasars.
However, large scale structure formation with magnetic field amplification and cosmic ray deflection simulations suggest smaller values no larger than $\BIGMF^0 = 2 \times 10^{-12}$\,G \cite{dolag2005}
or $\BIGMF^0 \approx 10^{-11}$\,G in voids \cite{sigl2004}.
The morphology of the IGMF is not known either and the most simple assumption is a domain-like structure which is also adopted here.
The field strength is constant in each cell and only grows with cosmic expansion, i.e. $\BIGMF(z) = \BIGMF^0(1+z)^2$, but the orientation changes randomly from one cell to another.
Furthermore, the domain length is given by $\CIGMF$. 
As shown in Ref. \cite{wouters2012}, adopting a Kolmogorov-type turbulence spectrum instead of the simple domain structure has negligible effects on the results.
In principle, the same procedure is followed here as presented in Ref. \cite{deangelis2011}, with the exception that 
the assumption of a the strong mixing is dropped.

\begin{figure}[t]
 \centering
 \begin{tabular}{cc}
 \includegraphics[width = 0.48 \linewidth]{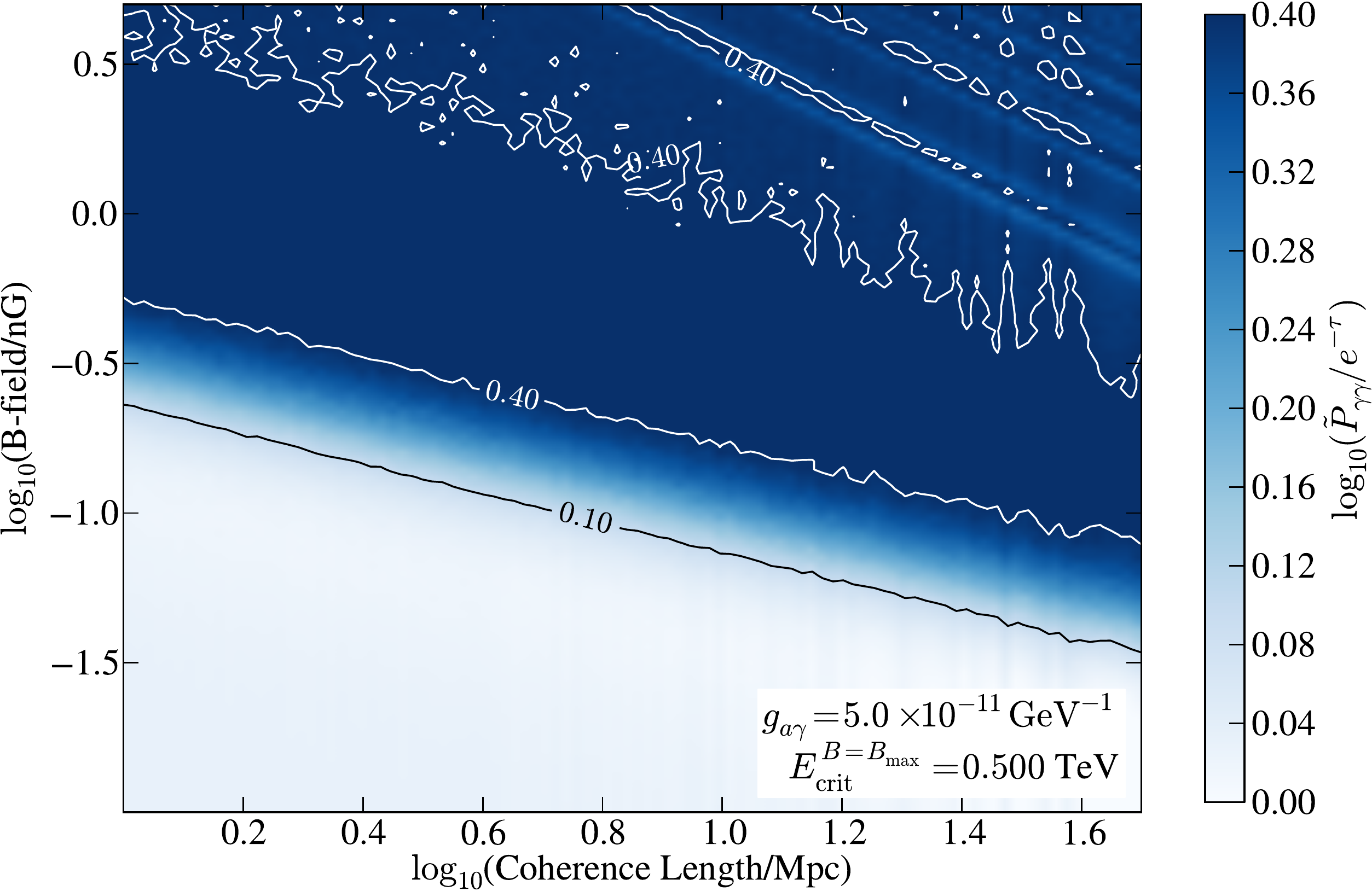}\hspace{0.5cm} &
 \includegraphics[width = 0.48 \linewidth]{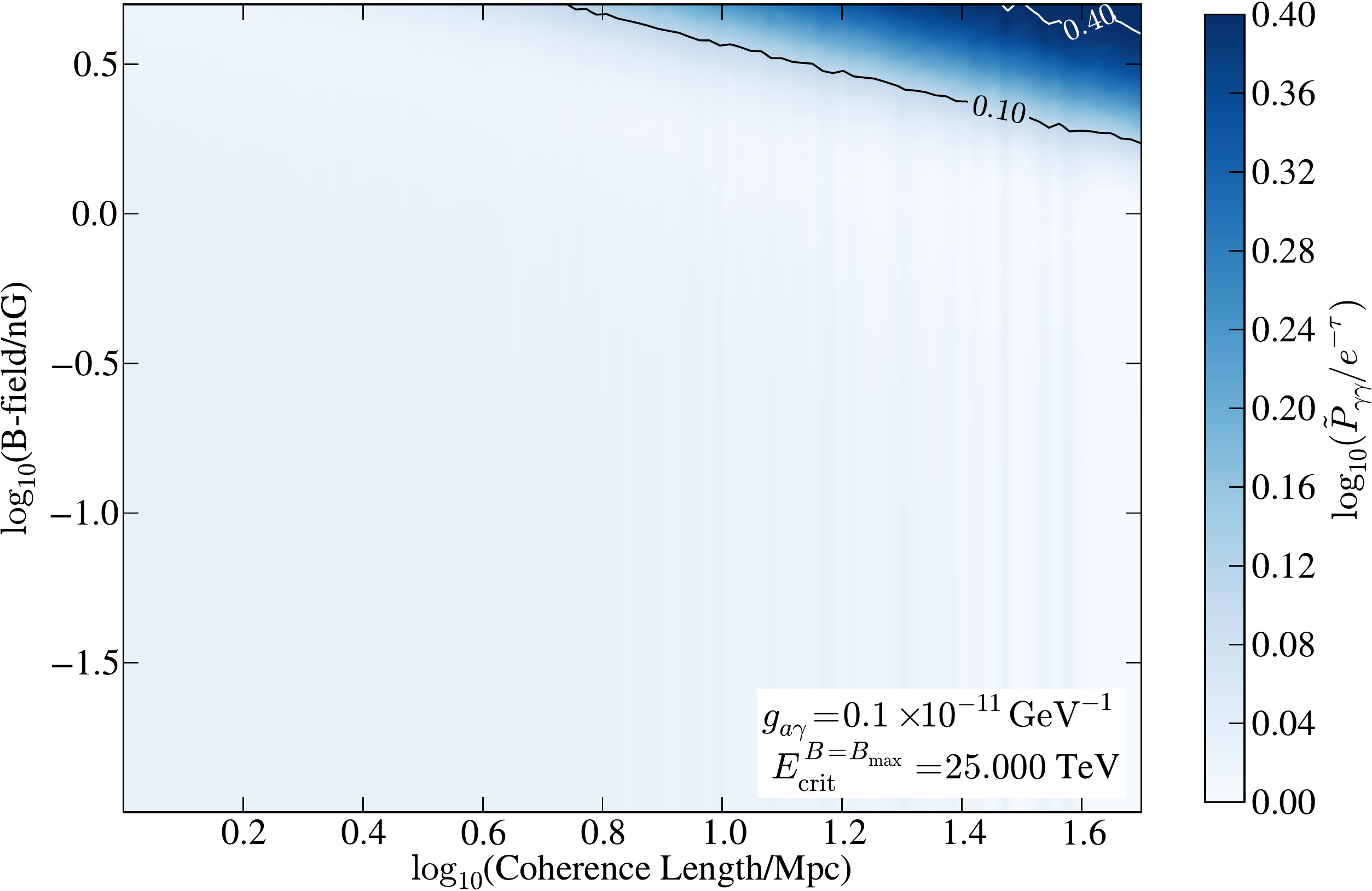} \\
 \includegraphics[width = 0.48 \linewidth]{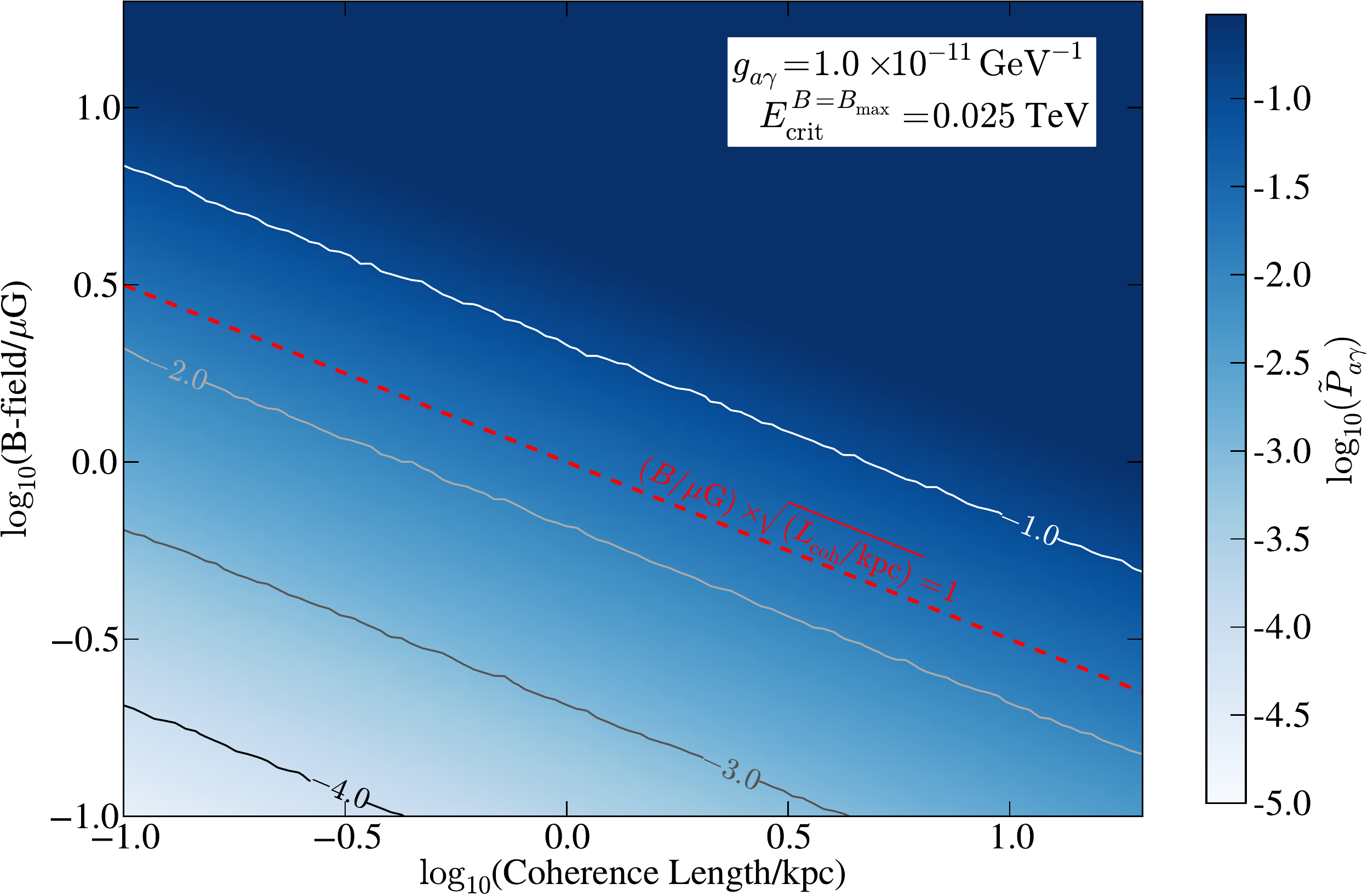}\hspace{0.5cm} & 
 \includegraphics[width = 0.48 \linewidth]{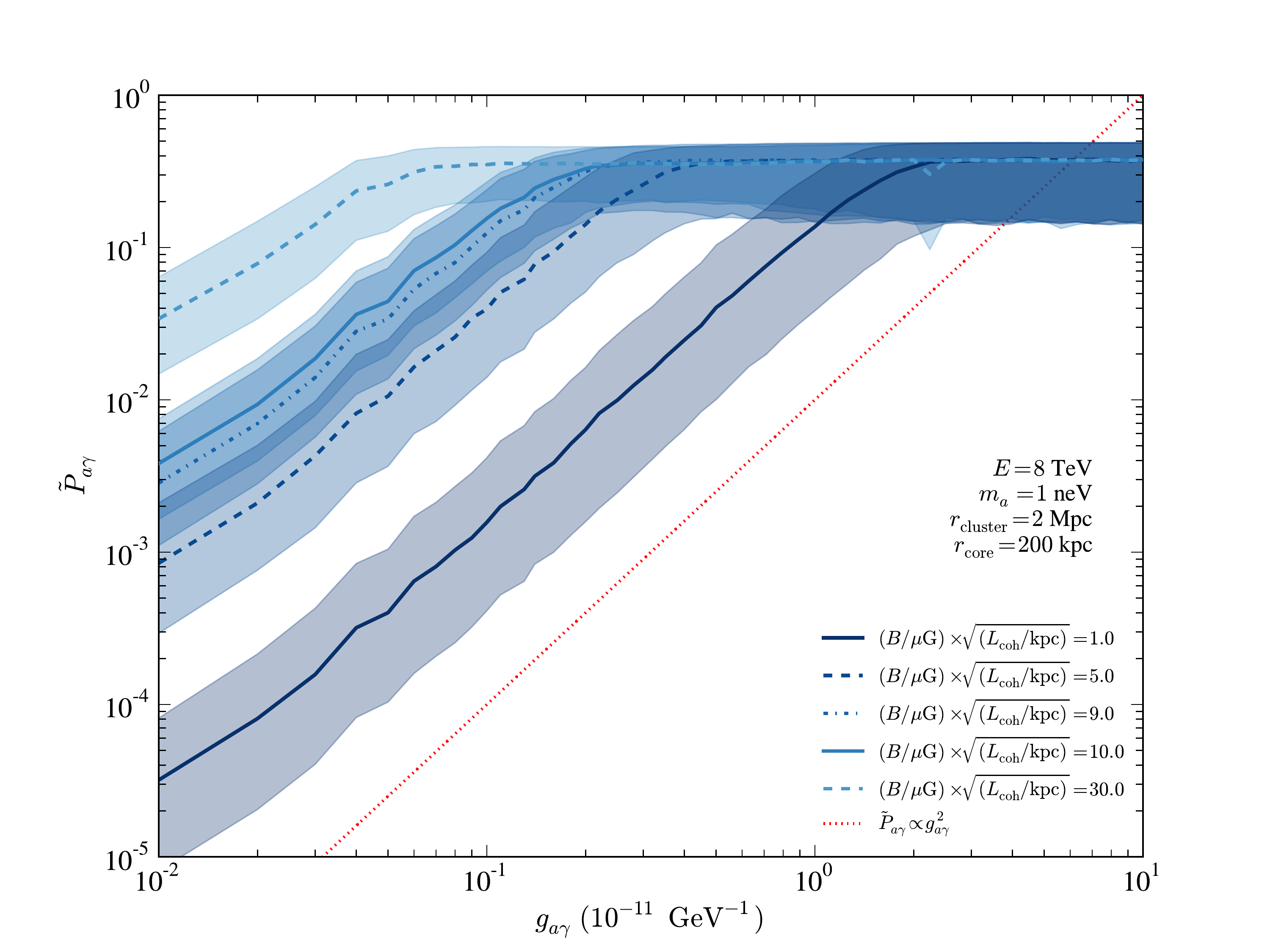} 
 \end{tabular}
 \caption{Parameter space scan in the $(\lambda_c, B_0)$ plane. Top row: \Pa-conversion in the IGMF.
  The color map displays the boost factor of the median of all simulated $\BIGMF$-field realizations; see Eq. (\ref{eq:boost}).
  The adopted values for the coupling (Left column: large couplings; Right column: small couplings) are displayed in the figure 
  together with the critical energy above which the conversion occurs in the strong mixing regime.
  Bottom row: Conversion in the ICMF.
  In the left panel, the color coding shows the fraction of the initial photon beam that is converted to ALPs (median over all realizations).
  The median of the conversion probability is constant for constant values of $B_0\times \sqrt{L_\mathrm{coh}}$, as indicated by the red dashed line.
  The bottom-right  panel displays the dependence of $\tilde{P}_{a\gamma}$ on the coupling $\gag$ for different values of $B_0 \times \sqrt{L_\mathrm{coh}}$.
  In this panel, 68\,\% of all $B$-field realizations for each $B_0\times \sqrt{L_\mathrm{coh}}$ value fall into the corresponding shaded regions.
  See text for further details.
 }
\label{fig:b-scan}
\end{figure}

A scan over a logarithmic grid with $100\times100$ pixels in the $(\CIGMF,\BIGMF^0)$ space is performed to determine the most optimistic magnetic field setup. 
For each grid point, the photon survival probability is calculated with Eq. (\ref{eq:phot-surv})
for 5000 realizations of the orientation of $\BIGMF$ 
for a fixed source distance $z = 0.536$, energy $E = 0.574$\,TeV (this combination of $z$ and $E$ corresponds to an optical depth of $\tau = 4$ with the \KD),
an ALP mass $\ma = 0.1$\,neV, and two different values of the coupling.
Only the conversion in the intergalactic magnetic field with absorption due to the EBL of the \KD~is taken into account.
The impact of the \pa~conversions is quantified with the boost factor $\mathcal{B}$, defined by
\begin{equation}
 \mathcal{B} = \medphotsurv / \exp(-\tau),
 \label{eq:boost}
\end{equation}
where $\medphotsurv$ is the median of the distribution of photon survival probabilities
\footnote{The median is preferred over the mean value since the distribution of $P_{\gamma\gamma}$ is highly skewed \cite[see][]{horns2012ICM}.}.
The result is shown for two different values of the \pa-coupling in the top row of Fig. \ref{fig:b-scan}.
As one would naively expect, for a large coupling of $\gag = 5 \times 10^{-11}\mathrm{GeV}^{-1}$ (top-right panel) the boost factor increases with 
increasing $\BIGMF^0$ and increasing $\CIGMF$ up to a value of $10^{0.4} \approx 2.5$.
For even higher values, the boost factor starts to decrease again and shows an oscillatory behavior.
This feature was already observed in Ref. \cite{deangelis2011}: if the conversion probability becomes too high, 
the photon fraction in the beam is large
at all times; but at the same time, the photon flux is attenuated by the interaction with the EBL.
As a consequence, $\mathcal{B}$ declines, and one is tempted to choose the values of $\CIGMF$ and $\BIGMF^0$ from within the $0.4$ contour. 
The situation changes, however, if $\gag$ is decreased (top-left panel) by more than an order of magnitude to $10^{-12}\mathrm{GeV}^{-1}$.
The entire region of $\mathcal{B} > 0$ is shifted towards higher values in the $(\CIGMF,\BIGMF^0)$ plane.
Without any \emph{a priori} assumption about values of the ALP mass and coupling, it is thus advisable to select the maximum values 
of $\CIGMF$ and $\BIGMF^0$ that are allowed by observations and it is settled for $\CIGMF \mathbf{= 50}$\,Mpc
and $\BIGMF^0 = 5$\,nG.
For the thermal electron density in the intergalactic medium, a typical value of $n_\mathrm{el,\,IGM} = 10^{-7}\,\mathrm{cm}^{-3}$ is adopted \cite{wmap7obs}.

\subsection{Intracluster magnetic fields}
In contrast to intergalactic magnetic fields, the existence of intracluster magnetic fields is well established. 
Synchrotron emission of the intracluster medium together with Faraday rotation measurements at radio frequencies
have led to the common picture that turbulent magnetic fields of the order of $\mathcal{O}(\mu\mathrm{G})$ fill the cluster volume
(see e.g. Refs. \cite[][]{govoni2004,feretti2012} for reviews and typical values of the model parameters used below).
The turbulence is usually described 
with a Kolmogorov-type spectrum, or with the simpler cell-like structure which is again used here (as in Ref. \cite[][]{horns2012ICM}).
There is evidence that the magnetic field strength follows the radial profile of the thermal electron distribution  $n_\mathrm{el,\,ICM}$ in the cluster, 
\begin{equation}
 \BICMF(r) = \BICMF^0 \left(\fract{n_\mathrm{el,\,ICM}(r)}{n_\mathrm{el,\,ICM}^0}\right)^\eta,
 \label{eq:icmf}
\end{equation}
with typical values $0.5 \lesssim \eta \lesssim 1$ and central magnetic fields up to $\sim 10\,\mu$G in the most massive clusters.
The thermal electron density is described by 
\begin{equation}
 n_\mathrm{el,\,ICM}(r) = n_\mathrm{el,\,ICM}^{0}\left(1 + \fract{r}{r_\mathrm{core}}\right)^{-3\beta / 2},
 \label{eq:nicm}
\end{equation}
with characteristic values of $\beta = 2 / 3$ and $r_\mathrm{core} = 200$\,kpc.
The coherence length is usually assumed to be comparable to galactic scales of the order of 10\,kpc.
As before, a grid scan over the $(\CICMF,\BICMF^0)$ plane is performed in order to determine the parameters that maximize the \pa~conversions.
A cluster with a radius of 2\,Mpc is assumed together with $\eta = 0.5$.
Instead of the boost factor, the fraction of ALPs $P_{a\gamma}$ in the final state [i.e., $\rho_\mathrm{final} = \rho_{33} = (0,0,1)$] is shown in the bottom row of Fig. \ref{fig:b-scan}
for an initially unpolarized pure photon beam, $P_{a\gamma} = \mathrm{Tr}(\rho_{33}\mathcal{T}\rho_\mathrm{unpol}\mathcal{T}^\dagger)$.
Again, 5000 $\BICMF$-field realizations are simulated, and the median $\tilde{P}_{a\gamma}$ is computed.
The more ALPs leave the cluster the stronger the effect will be on the VHE spectra because more ALPs can convert back into photons in the GMF and enhance the observed flux.
The bottom-left panel of Fig. \ref{fig:b-scan} clearly shows that more ALPs are produced for stronger magnetic fields and longer coherence lengths for a \pa~coupling strength of $\gag = 10^{-11}\mathrm{GeV}^{-1}$.
Interestingly, the conversion probability is constant for constant values of $B_0 \times \sqrt{L_\mathrm{coh}}$ (red dashed line), and
$\tilde{P_{a\gamma}}$ increases quadratically with growing coupling strength 
until the maximum probability of $\sim 1 / 3$ is reached (bottom-right panel of Fig. \ref{fig:b-scan}).
Not surprisingly, the maximum is reached for smaller couplings for larger values of $B_0 \times \sqrt{L_\mathrm{coh}}$.
Thus, for an optimistic scenario, a central ICMF value of $B_0 = 10\,\mu$G with a coherence length of 10\,kpc is chosen
and it is assumed that the VHE-emitting AGN is located at the center of a galaxy cluster.
The core thermal electron density is taken to be $n_\mathrm{el,\,ICM}^0 = 10^{-2}\,\mathrm{cm}^{-3}$.

\subsection{Magnetic field scenarios}
Now that the most optimistic values for the different magnetic fields are identified,
four scenarios are presented for which the effect of \pa-oscillations on VHE-\gray~spectra will be investigated.
In all four configurations, the conversion in the GMF is included.
\begin{enumerate}
 \item In the first scenario, called \scenmax~hereafter, no specific environment is assumed for the ALP production and  only the conversion in the GMF is included.
Instead, an initial
 beam polarization
 $\rho_\mathrm{init} = 1/3\, \mathrm{diag}(e^{-\tau}, e^{-\tau}, 1)$ is considered.
 This situation corresponds to a maximal mixing in some turbulent magnetic field inside or around the source and a subsequent attenuation of the photon fraction of the beam.
 In this general scheme, one is not forced to apply some sort of averaging over the many possible orientations of the random magnetic field.
 \item In a second configuration, named \scenICM, it is optimistically assumed that \emph{all} VHE \gray-emitting AGN are located at the center of galaxy clusters of a 2\,Mpc radius.
 The magnetic field changes over the distance from the cluster core as in Eq. (\ref{eq:icmf}) with a central magnetic field of $\BICMF^0 = 10\,\mu\mathrm{G}$
 and a coherence length of $\CICMF = 10$\,kpc.
 Any conversion in the intergalactic magnetic field is neglected, as well as any attenuation of the photon flux by local radiation fields inside the galaxy cluster. 
 Upon exit of the galaxy cluster, the photon beam will be attenuated by the interaction with the EBL whereas the ALP fraction propagates unhampered over the entire distance to the Milky Way.
 In the GMF, ALPs and photons can again convert into each other.
 Apart from the ICMF that drops radially with distance from the cluster core, this is the same setup as investigated in Ref. \cite{horns2012ICM}.
 \item Thirdly, it will be assumed that no AGNs are affected by the \pa-conversion inside a galaxy cluster but, on the other hand, 
the intergalactic field will be taken to its most optimistic values, i.e., $\BIGMF^0 = 5$\,nG and $\CIGMF = 50$\,Mpc. This setup is labeled \scenIGMF
 and is basically the same as that considered in, e.g., Ref. \cite{deangelis2011} apart from the complete energy-dependent treatment applied here.
 \item Finally, a set of more conservative model parameters is chosen to study both the conversion in the IGMF and ICMF.
 The parameters are conservative in the sense that they are not as close to the observational bounds as in the optimistic scenarios introduced above.
 Only the AGN listed in Tab. 1 of \cite{horns2012ICM} are assumed to be located inside a galaxy cluster. As their position relative to the cluster core in unknown,
 a constant ICMF of 1\,$\mu$G is assumed.
 Furthermore, a value of $r_\mathrm{cluster} = 2/3$\,Mpc is adopted as the distance that photons propagate through the intra-cluster medium 
 \footnote{This value is motivated by the following reasoning: assume that an AGN is placed randomly inside a sphere with a radius of 2\,Mpc, one can compute the distance to the edge of the sphere in a certain direction.
 If this is repeated a large number of times, the median distance that a photon travels through the sphere is found to be approximately 2/3.}.
 The value of the IGMF is motivated by simulations of large scale structure formations \cite{sigl2004,dolag2005}.
 This framework will be called \scenfid. 
\end{enumerate}
All scenarios are analyzed with two EBL models, namely the model of Ref. \cite[][]{franceschini2008} (henceforth the \FRV) and the lower limit prediction of the \KD.
The optical depth of the former is additionally scaled by a factor of $\sim 1.3$, as suggested by recent studies of VHE-\gray~spectra \cite{hess2013ebl}.
These two EBL models more or less bracket the range of the EBL density allowed by lower limits from galaxy number counts mentioned in Sec. \ref{sec:intro} and upper limits derived from VHE-\gray~spectra \cite{meyer2012}.
Moreover, it was shown in Ref. \cite{meyer2012ppa} that these two models result in the highest significance of the pair production anomaly;
and it can be expected that comparably small photon-ALP couplings are able to reduce this tension significantly and, thus,
 to derive conservative lower limits
on the photon-ALP coupling.
The different scenarios and their corresponding model parameters are summarized in Table \ref{tab:modpars}.
The \pa-conversion inside the source is not explicitly taken into account here, but a possible contribution is accounted for in the \scenmax~scenario.

\begin{table}[t]
 \begin{small}
 \centering
 \caption{Model parameters for the different magnetic field scenarios.
 In frameworks including the conversion inside galaxy clusters, the beam is assumed to travel the distance $r_\mathrm{cluster}$ through the volume filled with a $B$-field.
 In the \scenICM~scenario, the $B$ field varies as in Eq. (\ref{eq:icmf}). All AGN are assumed to be located at the center of a cluster.
 In the \scenfid~case, the magnetic field and thermal electron density are 
 assumed to be constant throughout the cluster volume.
 Only AGN listed in Table 1 of Ref. \cite{horns2012ICM} are assumed to lie within galaxy clusters. See text for further details.}
 \label{tab:modpars}
 \begin{tabular}{l|ccc|cccccc}
  \hline
  \hline
  {} & \multicolumn{3}{|c|}{IGMF} & \multicolumn{6}{c}{ICM}	\\
  Name & $\BIGMF^0$  & $\CIGMF$ & $n^0_\mathrm{el,\,IGM}$  & $\BICMF^0$  & $\CICMF$  & $r_\mathrm{cluster}$  & $n^0_\mathrm{el,\,ICM}$  &  $r_\mathrm{core}$ & $\eta$ \\
  {} &  (nG) &  (Mpc)& ($\times 10^{-7} \mathrm{cm}^{-3}$) & ($\mu$G) &  (kpc) &  (Mpc) &  ($\times10^{-3}\mathrm{cm}^{-3}$) &  (kpc) & {} \\
  \hline
  \scenmax & \multicolumn{9}{c}{Only conversion in GMF, but $\rho_\mathrm{init} = 1/3\,\mathrm{diag}(e^{-\tau},e^{-\tau}, 1)$} \\
  \hline
  \scenIGMF & 5 & 50 & 1 & $\ldots$ & $\ldots$ & $\ldots$ & $\ldots$ & $\ldots$ & $\ldots$  \\
  \scenICM & $\ldots$ & $\ldots$ & $\ldots$ & 10 & 10 & 2 & 10 & 200 & 0.5  \\
  \scenfid & 0.01 & 10 & 1 & 1 & 10 & 2/3 & 1 & $\ldots$ & $\ldots$  \\
  \hline
 \end{tabular}
 \end{small}
\end{table}

\section{Probing the Opacity with VHE gamma-ray spectra}
\label{sec:method}

With the framework to calculate the photon survival probability $P_{\gamma\gamma}$ introduced in the previous sections,
the observed VHE-\gray~spectra are corrected for absorption in the presence of ALPs.
As it is not assumed that the \pa-conversions occur in the strong mixing regime, 
$P_{\gamma\gamma}$ can show a strong oscillatory behavior. 
Therefore, an observed spectral point with a flux $\Phi^\mathrm{obs}_i$ over an energy bin $\Delta E_i$ with central energy $E_i$ is corrected 
with an average transfer function, 
\begin{equation}
 \langle P_{\gamma\gamma} \rangle_i = \frac{1}{\Delta E_i}\int_{\Delta E_i} \mathrm{d} E\, P_{\gamma\gamma}(E),
\end{equation}
so that the absorption corrected flux $\Phi_i$ is obtained by
\begin{equation}
 \Phi_i = \langle P_{\gamma\gamma} \rangle_i^{-1}\, \Phi_i^\mathrm{obs}.
\end{equation}
In practice, the photon survival probability is calculated for 40 energies for each source and linearly interpolated in $\log_{10}(E)$ and $\log_{10}(P_{\gamma\gamma})$.
This has been cross-checked for one $(\ma, \gag)$ pair with 100 energies and the results are found to be compatible if only 40 energies are used.

The same technique as put forward in Appendix B of \cite{horns2012} is used here to quantify the significance of the pair production anomaly in the presence of ALPs.
Each spectrum $j$ with data points that correspond to $\tau(z,E_i) \geqslant 2$, i.e. the optical thick regime, is fitted with an analytical function $f_j(E)$.
A list with all considered spectra that fulfill this criterion is shown in Table \ref{tab:spectra}.
The function $f_j(E)$ is either a power law, or, in case the fit probability is $p_\mathrm{fit} < 0.05$, a logarithmic parabola,
\begin{equation}
 f_j(E) = \begin{cases}
           N_0 (E / E_0)^{-\Gamma}, & p_\mathrm{fit}\geqslant 0.05, \\
	   N_0 (E / E_0)^{-(\Gamma  + \beta_c \ln(E / E_0))}, & \mathrm{otherwise},
          \end{cases}
\label{eq:fitfunc}
\end{equation}
with a flux and energy normalization $N_0$ and $E_0$, respectively; a power-law index $\Gamma$; and a curvature $\beta_c$.
For each data point in the optical thick regime, the residual is calculated,
\begin{equation}
 \chi_{ij} = \frac{\Phi_i - f_j(E_i)}{\sigma_i},
\end{equation}
which is normalized to the statistical measurement uncertainty $\sigma_i$ on the flux $\Phi_i$ (68\,\% confidence).
Under the hypothesis that $P_{\gamma\gamma}$ gives a correct prediction of the opacity of the Universe to VHE \gray s,
the residuals in the optical thick regime should follow a Gaussian distribution with zero mean.
This conjecture is checked with the $t$ test, for which the variable 
\begin{equation}
 t = \frac{\bar{\chi}}{\sqrt{\sigma_\chi / N_\chi}},\label{Eq:t}
\end{equation}
with $\bar{\chi}$ the mean and $\sigma_\chi$ the variance of the residual distribution that contains $N_\chi$ data points,
 follows a $t$ distribution from which the significance (one-sided confidence interval) can be calculated.
This method to quantify the accordance between model and data has several advantages. Firstly, the functions to parametrize the spectra do not 
depend on any particular blazar emission model, as no constraints on the photon index $\Gamma$ nor on the curvature $\beta_c$ are made during the fit.
Most spectra are adequately described by these functions, as shown in the Appendix.
Secondly, no extrapolation from the optical thin to the optical thick regime is required, and the statistical uncertainties of the measurement enter
the significance test self-consistently. 

Without the contribution of ALPs, one finds for the spectra listed in Table \ref{tab:spectra} 
 a significance of $5.1 \times 10^{-6} \approx 4.4\,\sigma$ for the \KD~and 
 $2.1\times10^{-4} \approx 3.5\,\sigma$ for the \FRV~scaled by 1.3 that the models do not describe the data.
It might come as a surprise that the scaled \FRV~gives a lower significance than the minimal attenuation model of Ref. \cite{kneiske2010}.
The reason for this is that more data points migrate into the optical thick regime as the EBL density increases.
This leads to an overall distribution closer to a zero mean
and shows the limitation of the method:
as long as the overall fit to all spectra is acceptable, the entire residual distribution must scatter around zero
(see also Ref. \cite[][]{meyer2012ppa}).

\begin{table}[t]
\caption{List of VHE-\gray~spectra included in the analysis. The table shows the redshift of the source, the IACT experiment that measured it,
the energy range covered by the spectrum and the number of data points in the optical thick regime for the optical depth given by the \KD~and
by the scaled version of the \FRV.
}
\label{tab:spectra}
 \centering
 \begin{tabular}{llcccccc}
  \hline
  \hline
  \multirow{2}{*}{$j$} & \multirow{2}{*}{Source} & \multirow{2}{*}{Redshift} & \multirow{2}{*}{Experiment} & Energy range &  $N_{\tau \geqslant 2}$ & $N_{\tau \geqslant 2}$ & \multirow{2}{*}{Reference}\\
  {}& 	{}  & {} & {} & (TeV) & ($\tau = 1 \times\,\tau_\mathrm{KD}$) &($\tau = 1.3 \times\,\tau_\mathrm{FRV}$) &{}\\
  \hline
  1 & Mrk\,421 & 0.031 & HEGRA &  0.82 -- 13.59 & 0 & 1 & \cite{mkn421hegra2002} \\ 
  2 & Mrk\,421 & 0.031 & HEGRA &  0.82 -- 13.59 & 0 & 1 & \cite{mkn421hegra2002} \\ 
  3 & Mrk\,421 & 0.031 & \hess &  1.12 -- 17.44 & 0 & 2 & \cite{mkn421hess2005} \\ 
  4 & Mrk\,421 & 0.031 & \hess &  1.75 -- 23.10 & 1 & 4 & \cite{mkn421hess2010} \\ 
  5 & Mrk\,501\footnotemark[1] & 0.034 & HEGRA &0.56 -- 21.45 & 1 & 3 & \cite{mkn501hegra1999} \\
  6 & 1ES\,1950+650 & 0.048 & HEGRA & 1.59 -- 10.00 & 0 & 1 & \cite{1es1959hegra2003} \\
  7 & 1ES\,1950+650 & 0.048 & HEGRA & 1.52 -- 10.94 & 0 & 1 & \cite{1es1959hegra2003} \\
  8 & PKS\,2155-304\footnotemark[1] & 0.116 & \hess & 0.23 -- 2.28 & 0 & 2 & \cite{pks2155hess2005a} \\
  9 & PKS\,2155-304\footnotemark[1] & 0.116 & \hess & 0.23 -- 3.11 & 0 & 3 & \cite{pks2155hess2005b} \\
  10 & PKS\,2155-304\footnotemark[1] & 0.116 & \hess & 0.22 -- 4.72 & 0 & 6 & \cite{pks2155hess2007} \\
  11 & PKS\,2155-304\footnotemark[1] & 0.116 & \hess & 0.25 -- 3.20 & 0 & 2 & \cite{pks2155hess2009} \\
  12 & RGB\,J0710+591& 0.125 & VERITAS&0.42 -- 3.65 & 0 & 2 & \cite{rgbj0710veritas2010} \\
  \multirow{2}{*}{13} & \multirow{2}{*}{H\,1426+428}&\multirow{2}{*}{0.13} &HEGRA, & \multirow{2}{*}{0.25 -- 10.12} & \multirow{2}{*}{2} &\multirow{2}{*}{5} & \multirow{2}{*}{\cite{h1426combined2003}} \\
  {} & {} &{}&CAT, WHIPPLE & {} & {} & {} & {} \\
  14 & 1ES\,0229-200& 0.140 & \hess & 0.60 -- 11.45 & 3 & & \cite{1es0229hess2007}\\
  15 & H\,2356-309 & 0.165 & \hess & 0.18 - 0.92 & 0 & 1 & \cite{1es1101h2356hess2006}\\
  16 & H\,2356-309 & 0.165 & \hess & 0.22 - 0.91 & 0 & 1 & \cite{h2356hess2006}\\
  17 & H\,2356-309 & 0.165 & \hess & 0.23 - 1.71 & 0 & 1 & \cite{h2356hess2010}\\
  18 & 1ES\,1218+304 & 0.182 & VERITAS & 0.19 -- 1.48 & 0 & 3 & \cite{1es1218veritas2009}\\
  19 & 1ES\,1101-232\footnotemark[1] & 0.186 & \hess & 0.18 -- 2.92 & 3 & 7 & \cite{1es1101h2356hess2006}\\
  20 & 1ES\,0347-121 & 0.188 & \hess & 0.30 -- 3.03 & 2 & 4 & \cite{1es0347hess2007}\\
  21 & RBS\,0413\footnotemark[2] & 0.190 & VERITAS& 0.23 -- 0.61 & 0& 1 & \cite{rbs0413veritas2012}\\ 
  22 & 1ES\,0414+009\footnotemark[1] & 0.287 & \hess & 0.17 -- 1.13 & 2 & 3  & \cite{1es0414hess2012}\\
  23 & 1ES\,0414+009\footnotemark[1]${}^\mathrm{,}$\footnotemark[2]& 0.287 & VERITAS & 0.23 -- 0.61 & 0 & 1 & \cite{1es0414veritas2012}\\
  24 & PKS\,1222+21  & 0.432 & MAGIC & 0.08 -- 0.35 & 0 & 1 & \cite{pks1222magic2011} \\
  25 & 3C\,279\footnotemark[2] & 0.536 & MAGIC & 0.15 -- 0.35 & 1 & 1 & \cite{3c279magic2011}\\
  26 & 3C\,279 & 0.536 & MAGIC & 0.08 -- 0.48 & 1 & 2 & \cite{3c279magic2008}\\
\hline
 \end{tabular}
\footnotetext[1]{Assumed to be located in a galaxy cluster in the \scenfid~scenario \cite[see][Tab. 1]{horns2012ICM}.}
\footnotetext[2]{Not included in \cite{horns2012}.}
\end{table}

In the following discussion, ALPs are included in the correction of the observed spectrum. 
For this purpose, the transition probability for all four scenarios and the two different EBL models is calculated separately for each source 
listed in Table \ref{tab:spectra}. 
This is necessary because each AGN has a different redshift (important for the attenuation) and a different position in the sky (influencing the conversion in the GMF).
Furthermore, $P_{\gamma\gamma}$ is computed over a grid of equally spaced values in the $(\log_{10}(\ma),\log_{10}(\gag))$ space.
For the coupling constant, the range $10^{-13}\,\mathrm{GeV}^{-1} \leqslant \gag \leqslant 10^{-10}\,\mathrm{GeV}^{-1}$ is chosen for all magnetic field frameworks.
The upper bound is motivated by the bound set by the CAST experiment of $\gag < 8.8\times10^{-11}\,\mathrm{GeV}^{-1}$ \cite{cast2007}
while for the lower bound the contribution of ALPs is expected to become negligible. 
On the other hand, the range of the tested ALP masses differs in the different scenarios.
It is determined by the critical energy given in Eq. (\ref{eq:Ecrit}) that should span an interval that includes the minimum and maximum energies of the VHE spectrum sample in 
Table \ref{tab:spectra}, namely 0.08\,TeV and 23.1\,TeV.
The different magnetic fields and thermal electron densities result in different mass ranges.
 For the \scenmax- and \scenICM~configurations, a mass range of $1\,\mathrm{neV} \leqslant \ma \leqslant 10^3\,\mathrm{neV}$ is chosen; whereas for the \scenIGMF~setup,
the smaller values of $\BIGMF$ and the ambient density lead to a shift in the mass to $10^{-1.5}\,\mathrm{neV} \leqslant \ma \leqslant10^{1.5}\,\mathrm{neV}$.
In the combined \scenfid~scenario, it is settled for the intermediate range $10^{-0.5}\,\mathrm{neV} \leqslant \ma \leqslant10^{2.5}\,\mathrm{neV}$.
A resolution of the grid of $32 \times 32 = 1024$ points is selected in the particular ranges of $(\log_{10}(\ma),\log_{10}(\gag))$.

A complication is introduced by the random magnetic fields in the scenarios apart from the \scenmax~case.
Since the exact orientation of the IGMF and ICMF in each domain is unknown, a large number $N_\mathrm{sim}$ of simulated random realizations is required.
Here, $N_\mathrm{sim}$ will be set to 5000, and, therefore, for each $(\ma,\gag)$ pair one ends up with 5000 values for the significance level of the $t$ test, $p_t$.
One solution would be to compute the median (or mean) of the transfer function and afterwards calculate $p_t$. 
However, in the averaging process all information on the $p_t$ distribution is lost, and it is unclear if this certain value is statistically suitable to deduce a lower limit on $\gag$. 
Instead, the $p_t$ distribution is used to determine the $p_t$ value
 for which 95\,\% of all $B$-field realizations result in a worse compatibility of the particular framework with the data (i.e., those realizations that result in a smaller $p_t$ value).
This particular significance is henceforth denoted as $p_{95}$.
In summary, for each scenario, one now has one $p_{95}$ value for each grid point in the $(\ma,\gag)$ space. 
The lower limit on $\gag$ is then defined as the contour line for which $p_{95} = \LL$.
In this way, for $(\ma,\gag)$ values below this contour line, 95\,\% of all $B$-field realizations result in compatibility of less than 1\,\% with the data.
For the two EBL models used here, this corresponds to a decrease of the significance 
of the pair production anomaly by a factor of $5.1\times10^{-4}$ (\KD) and $2.1\times10^{-2}$ (scaled \FRV).

\section{Results}
\label{sec:results}

\begin{figure}[t]
 \centering
 \begin{tabular}{cc}
 \includegraphics[width = 0.48 \linewidth]{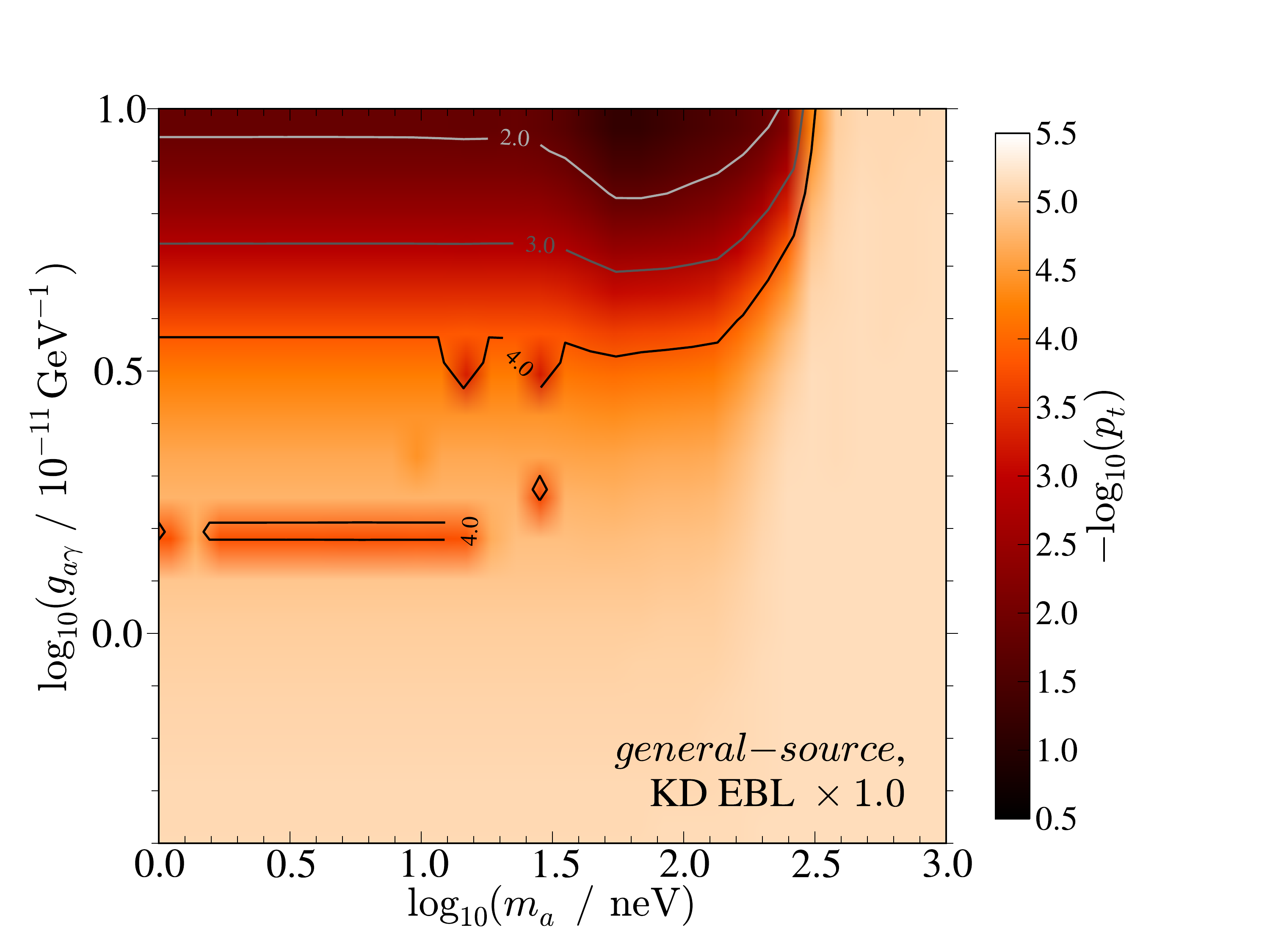}\hspace{0.2cm} & 
 \includegraphics[width = 0.48 \linewidth]{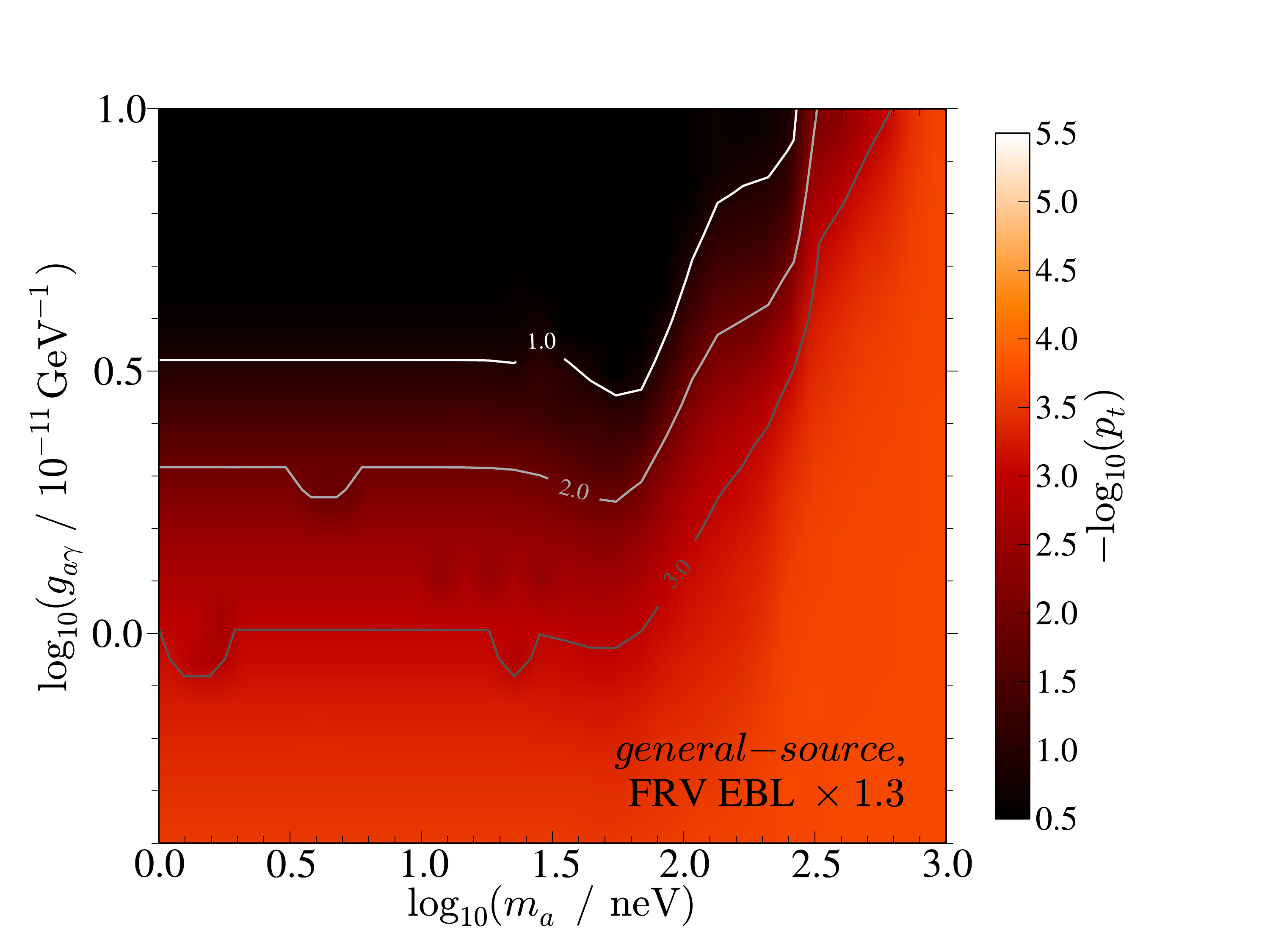} \\
 \includegraphics[width = 0.48 \linewidth]{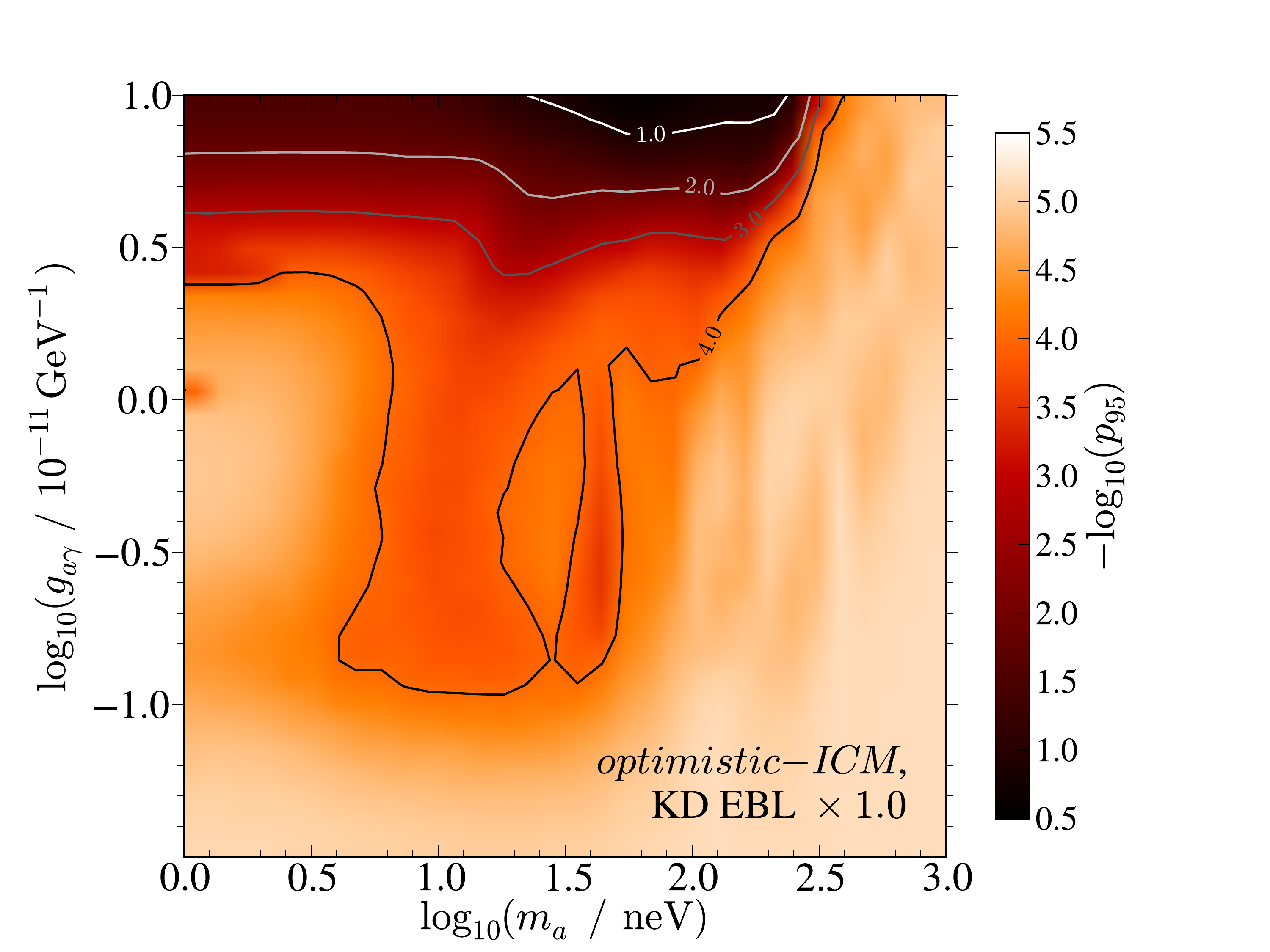}\hspace{0.2cm} & 
 \includegraphics[width = 0.48 \linewidth]{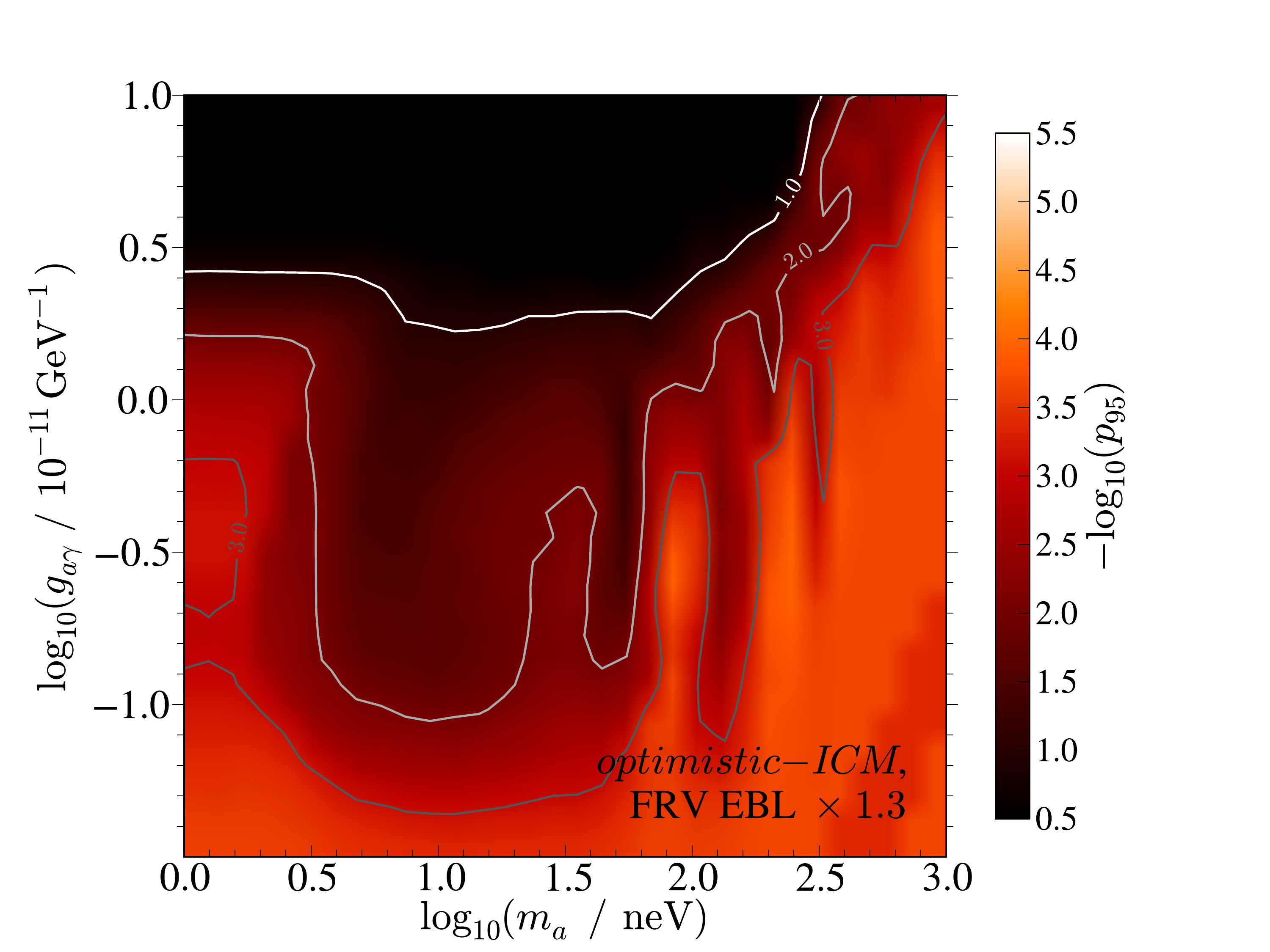} \\ 
 \end{tabular}
 \caption{Significance map for the \pa~conversion in the $(\ma,\gag)$ plane. 
Smaller values (brighter regions) indicate less accordance between the model and the data.
Upper panel: $p_t$ values for the \scenmax- scenario, shown as $-\log_{10}(p_t)$. 
Lower panel: $p_{95}$ values for the \scenICM~case. For each pixel, 5000 realizations of the random magnetic field are simulated and $p_{95}$ is determined from the resulting 5000 $p_t$ values (cf. Sec. \ref{sec:method}).
In the left column, the attenuation due to the interaction of VHE \gray s with the EBL is given by the \KD, whereas in the right column the \FRV~is utilized.
The maps are smoothed using a bilinear interpolation between the single $(\ma,\gag)$ pixels.}
 \label{fig:max-ICM}
\end{figure}

The results for the significance test introduced in the previous section are presented for each of the four scenarios developed in Sec. \ref{sec:Bfield}.
The upper panels of of Fig. \ref{fig:max-ICM} show the $p_t$ values for the \scenmax~configuration for the \KD~(top-right panel) and the \FRV~(top-left panel).
In this scenario, no random magnetic field is involved, and thus there is only one $p_t$ value for each pixel. 
The color coding and the contour lines show the $-\log_{10}(p_t)$ values and larger values of $p_t$ [smaller values of $-\log_{10}(p_t)$],
which represent a higher probability that the corresponding $t$ value is the result of a statistical fluctuation; i.e., a higher probability that the transfer function is in accordance with the data.
Clearly, the $p_t$ values increase with an increasing \pa-coupling.
The lower limit on $\gag$ ($p_t = \LL$ corresponds to the $-\log_{10}(p_t) = \logLL$ contour line) is 
$\sim7.8\times10^{-11}\,\mathrm{GeV}^{-1}$ for the \KD- and $\sim1.4\times10^{-11}\,\mathrm{GeV}^{-1}$  for the \FRV, respectively, 
in the regime where the mixing becomes independent of the ALP mass at $\ma \lesssim 15$\,neV.
This mass marks the onset of the strong mixing regime (SMR) for all spectra in the environment of the Milky Way.
For higher masses, the critical energy increases, and so does the number of spectral points outside the SMR.
Higher couplings of $\gag$ are necessary to compensate this effect and to retain a low level of the significance of the pair production anomaly.
Above $\ma \gtrsim 250$\,neV, the tested coupling does not lead to a reduction of the tension between the model and data in comparison to the no-ALPs case.
These observations are valid for both EBL models.

A similar overall behavior is found in the \scenICM~case(Fig. \ref{fig:max-ICM}, bottom-right panel: \KD; bottom-left panel: \FRV).
The color code now displays the $p_{95}$ values for the 5000 simulated realizations of the random ICMF in each pixel. 
Apart from the overall trend, peculiar regions are visible for the contour lines. 
In certain mass ranges, the lower limit contour for $\gag$ extends down to almost $\gag = 10^{-12}\,\mathrm{GeV}^{-1}$
using the \FRV.
These features are caused by the oscillatory behavior of $P_{\gamma\gamma}$ outside the SMR which affects the low-energy data points in the spectra.
These data points usually have the best count statistics, smallest error bars, and the strongest influence on the overall spectral fit.
The oscillations in the transfer function can lead to a correction that is strong in one energy bin but small in the adjacent bin.
As a result, the spectral fit is altered and leads to residuals in the optical thick regime that are closer to zero for certain $(\ma,\gag)$ pairs.

Thus, it is expected that these features will change if more VHE spectra are included in a future analysis.
Furthermore, the oscillations of the transfer function lead to a poor fit quality
for the spectra with the best overall statistics (Mrk\,421 \cite{mkn421hess2010}, Mrk\,501 \cite{mkn501hegra1999}, and PKS\,2155-304 \cite{pks2155hess2007})
and to a small overall fit probability (see the Appendix).
This will lead to a broadening of the residual distribution and a possible overestimation of the $p_{95}$ values closer to 1. 
These features should not be taken as a preferred parameter region for ALPs 
to explain the opacity of the Universe.

\begin{figure}[t]
 \centering
 \begin{tabular}{cc}
 \includegraphics[width = 0.48 \linewidth]{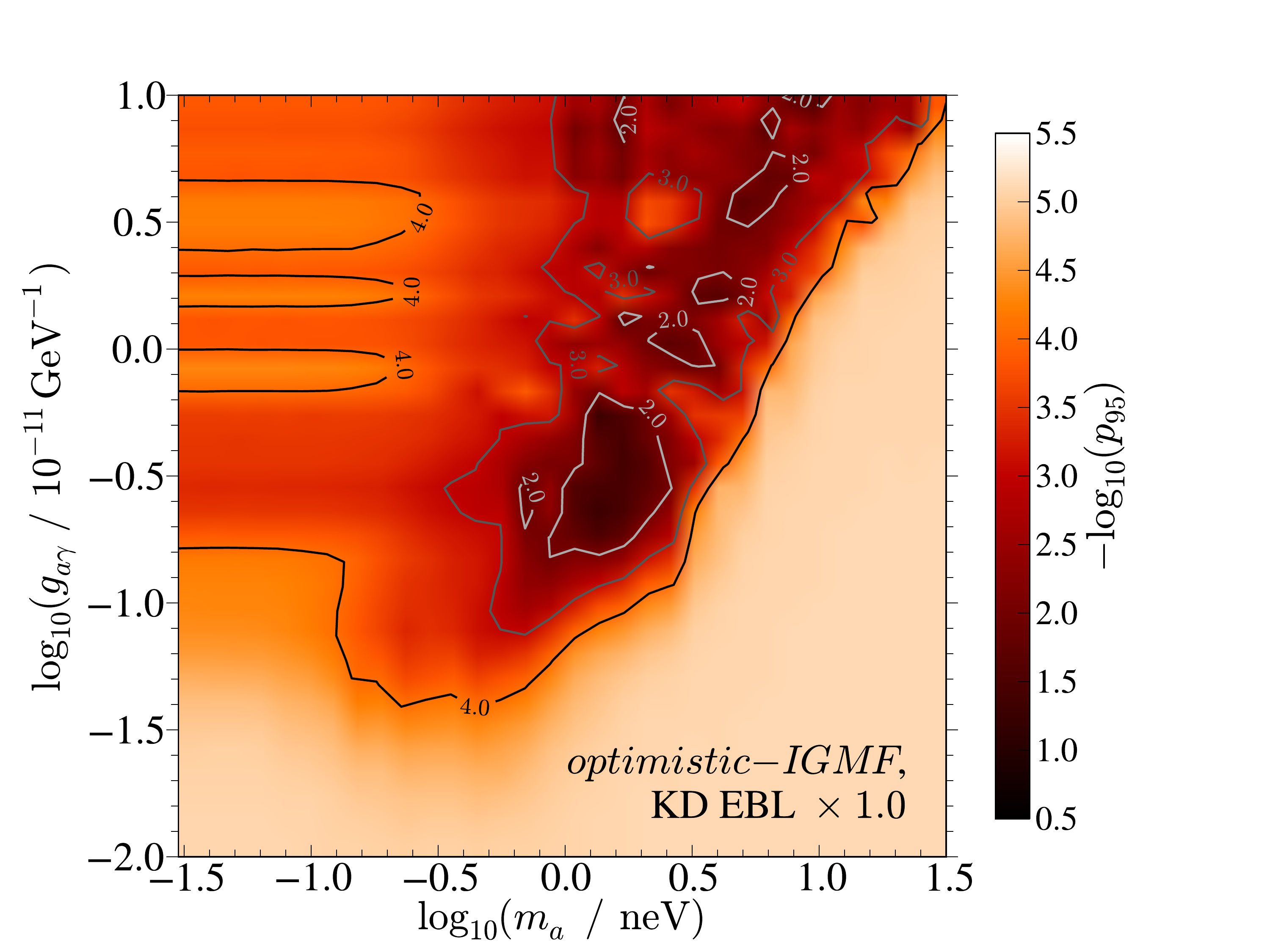}\hspace{0.2cm} & 
 \includegraphics[width = 0.48 \linewidth]{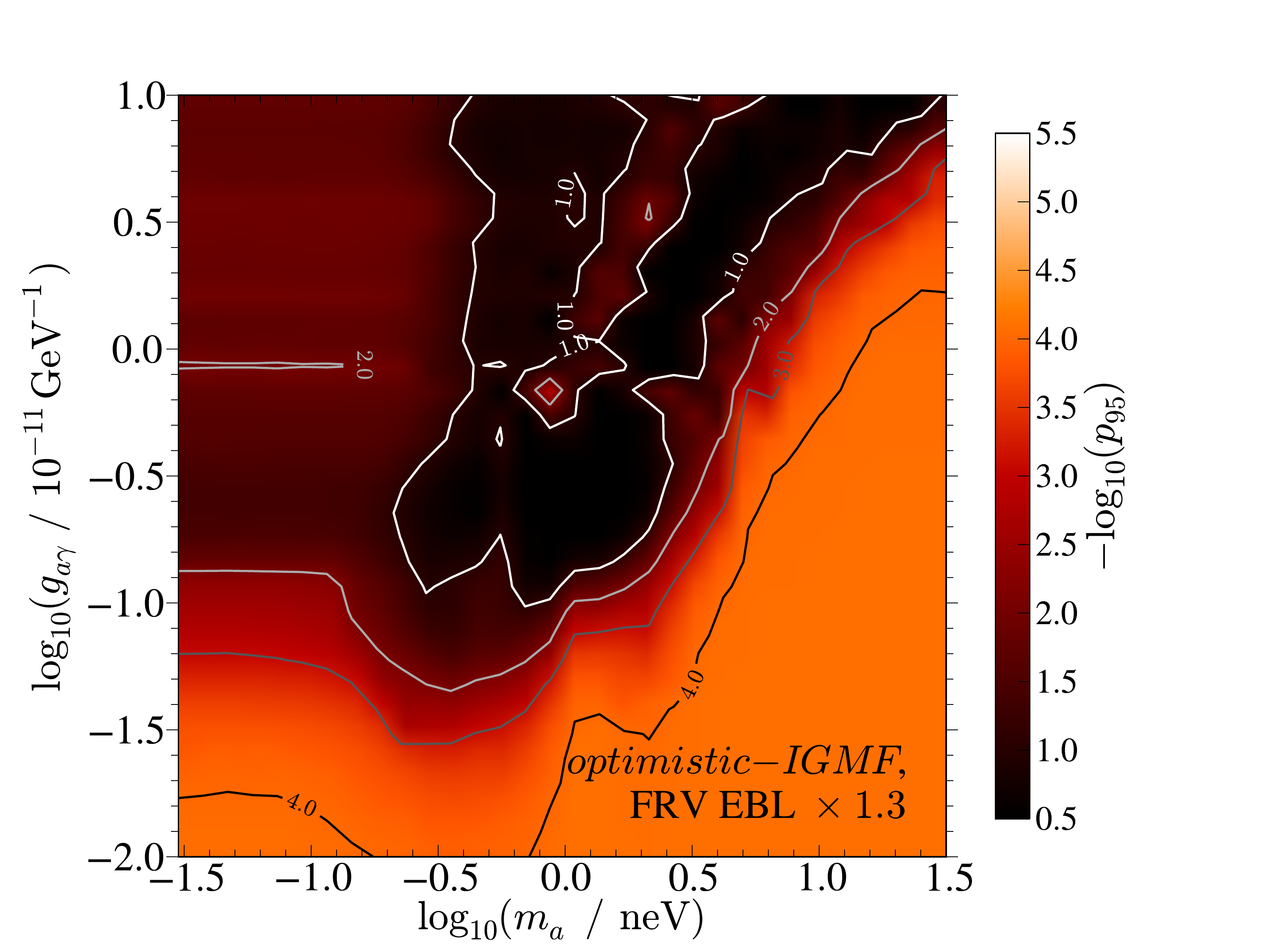} \\
 \includegraphics[width = 0.48 \linewidth]{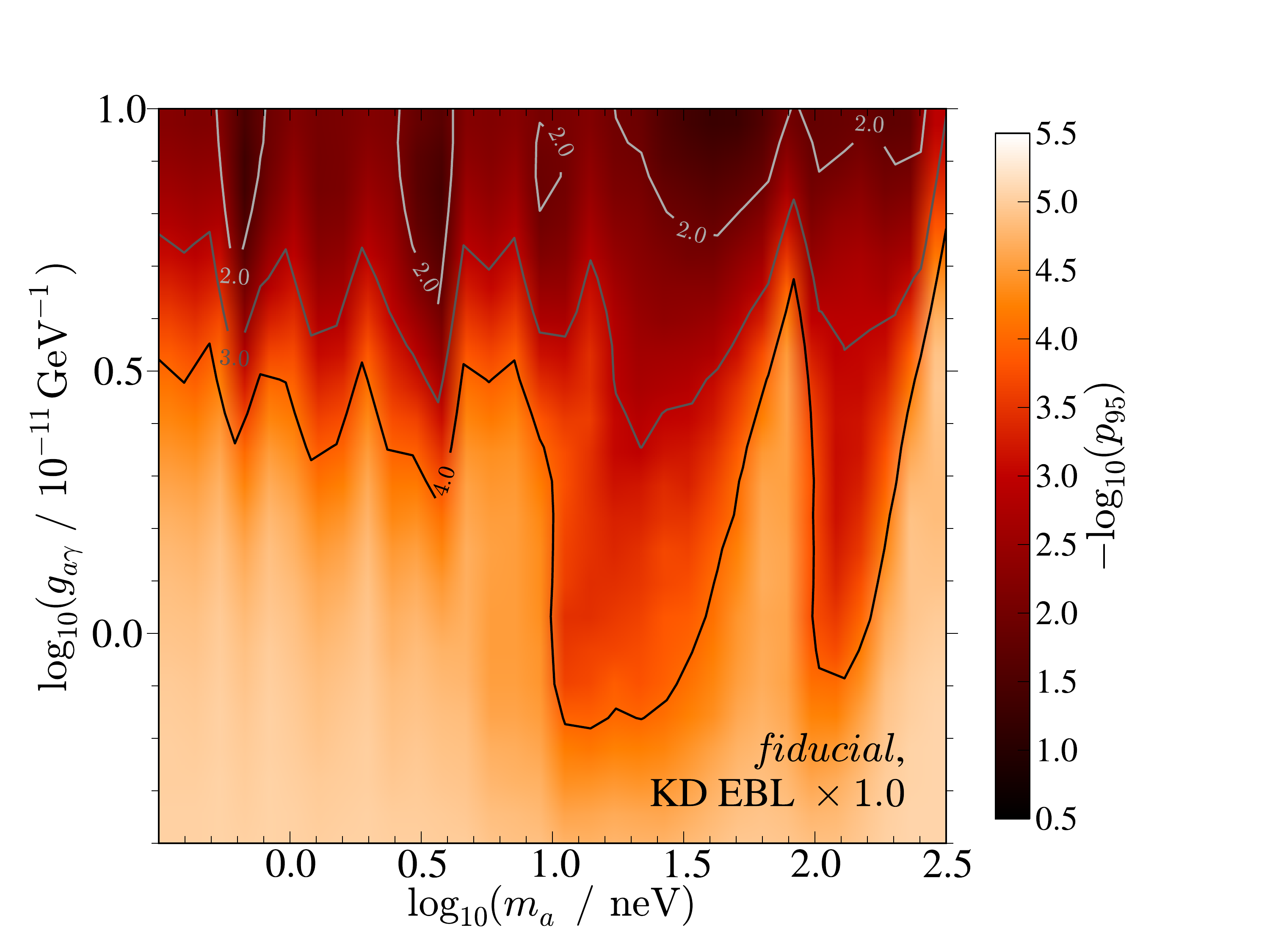}\hspace{0.2cm} & 
 \includegraphics[width = 0.48 \linewidth]{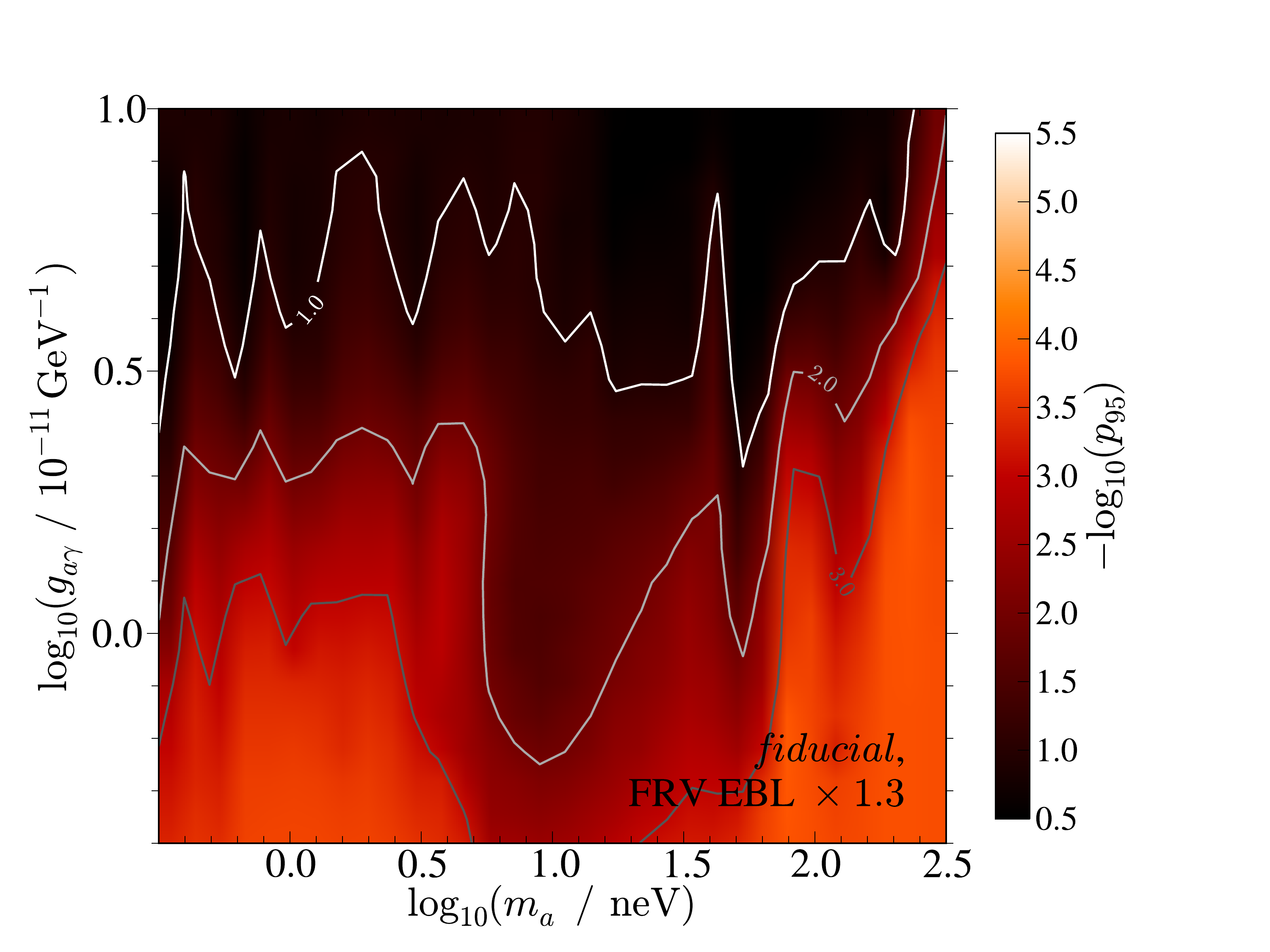}
 \end{tabular}
 \caption{Significance maps for the \pa-conversion. Same as the bottom row in Fig. \ref{fig:max-ICM}, but for the \scenIGMF-scenario (top row) and \scenfid-case (bottom row).}
\label{fig:IGMF-fid}
\end{figure}

In the \scenIGMF-scenario with the \KD, the only significant improvement over the no-ALP case is actually outside the SMR, as can be seen from Fig. \ref{fig:IGMF-fid} (top-right panel).
Note that the mass range in which the transition to the SMR occurs has now shifted to lower masses due to the smaller IGMF and ambient electron density compared to the intracluster case.
With the attenuation of the \FRV, the optimistic parameter choices for $\BIGMF^0$ and $\CIGMF$
 lead to a lower limit on $\gag$ -- as low as $\sim3\times10^{-13}\,\mathrm{GeV}^{-1}$ (top-left panel of Fig. \ref{fig:IGMF-fid}).

The bottom row of Fig. \ref{fig:IGMF-fid} displays the results for the more conservative parameter choice of the \scenfid-framework. 
In this scenario, one cannot strictly speak about a lower limit on $\gag$ as neither the values of the magnetic fields nor the values for the coherence lengths are set to their observationally allowed upper limits.
The $(\ma,\gag)$ pairs that result in $p_{95} \geqslant \LL$ can thus rather be seen as a preferred region in the parameter space 
if one tries to explain the opacity of the Universe with \pa-conversions.
One has to keep in mind, though, that the majority of simulated $B$-field realization results in smaller $p_t$ values.
Not surprisingly, one can conclude that the \pa~conversion in the IGMF is negligible, since the $p_{95} = \LL$ contour line does not extend to lower values of $\gag$ at $\ma \approx 1$\,neV,
as observed in the \scenIGMF-case.
Compared to the \scenICM-case, the lower limit contour line has shifted towards higher values in $\gag$ because a smaller number of AGNs is assumed to be located inside galaxy clusters. 
However, it has to be underlined that ALPs are still able to improve the accordance of the model with the data significantly.

From Figs. \ref{fig:max-ICM} and \ref{fig:IGMF-fid} it is obvious that in the \KD, higher values of the \pa-coupling are necessary to reduce the tension between model and data below the threshold of $p_{95} = 0.01$ (more stringent lower limits) than in the scaled \FRV. The reason for this is twofold: 
On the one hand, without ALPs,
the absorption correction in the \FRV~is larger for high optical depths, which leads to higher residuals in some spectra.
Lower \pa-couplings suffice in these cases to reduce the residuals. 
On the other hand, the significance of the pair production anomaly is lower in the scaled \FRV~to begin with (cf. Sec. \ref{sec:method}). 
Demanding the same decrease of significance as in the \FRV- without ALPs to the lower limit value ($2.1\times10^{-4}$ to 0.01)
in the \KD-case results in a significance value of 
$\sim 2.4\times10^{-4}$, close to the $p_t = 10^{-4}$ contour line. 
Especially in the \scenIGMF~and \scenfid~scenarios this line is in good agreement with the $p_t = 10^{-2}$ contour line in the \FRV~case.

VHE-\gray~spectra are subject to systematic uncertainties which can also affect the significance test used here.
The authors of Ref. \cite{horns2012} identified several sources of uncertainties  in the quantification of the significance of the pair production anomaly
such as a selection bias of VHE sources, the uncertainty of the overall energy scale of IACTs, and spillover effects in the highest energy bins due to the limited 
energy resolution of IACTs (the reader is referred to Ref. \cite{horns2012} for a detailed discussion). 
Including these effects leads in general to a reduction of the significance.
The strongest reduction, to 2.6\,$\sigma$, was found if the last energy bins of all spectra were excluded from the analysis and the energy points were simultaneously 
scaled by --15\,\% in energy (a conservative choice, as it was shown that a scaling of the order of 5\,\% is in better agreement with a cross correlation between IACTs and the \emph{Fermi}-LAT \cite{meyer2010}). This certainly poses a lower limit on the significance, as it seems unlikely that all VHE spectra are influenced by these systematics in the same way. 
Nevertheless, a lower limit of $p_{95} = \LL$ with the inclusion of ALPs would not help to significantly 
improve the accordance between model and data in this case of a marginal indication.
If ALPs were required at all, higher photon-ALP couplings would be necessary. 
However, the goal here is to set lower limits on $\gag$ if the pair production anomaly is not explained by invoking all systematic uncertainties on the VHE observations at once.

\begin{figure}[t]
 \centering
 \includegraphics[width = 0.98 \linewidth]{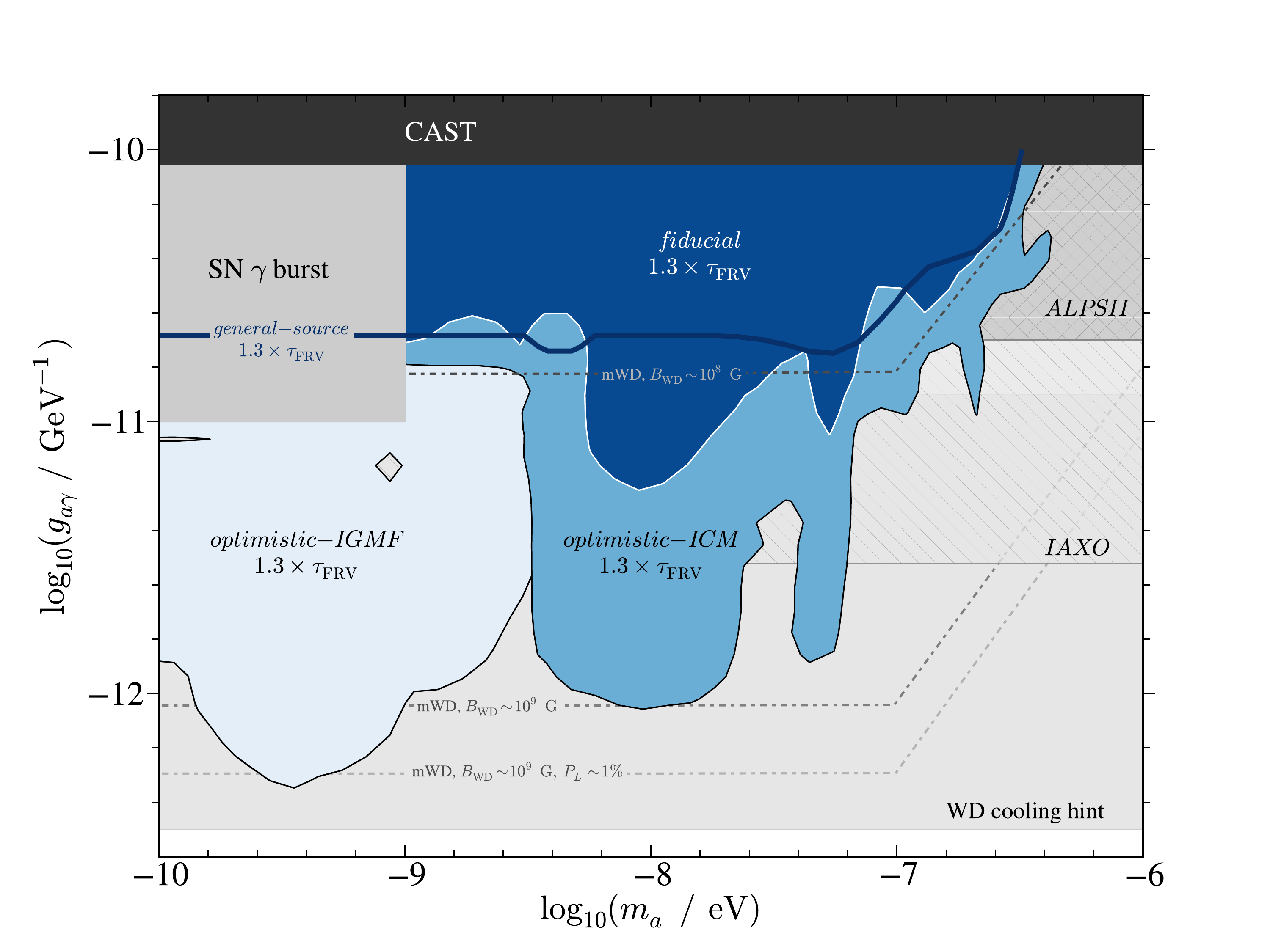}
 \caption{ALP parameter space with the lower limits on $\gag$ derived in this paper. The lower limits for the different scenarios are displayed as blue shaded regions, or
in the case of the \scenmax-scenario, as a dark blue solid line. They are only shown for the scaled \FRV, so that the optical depth is given by $\tau = 1.3\times\tau_\mathrm{FRV}$.
For comparison, upper limits, hints for theoretical preferred regions, and sensitivities of future experiments are also plotted. See text for further details.} 
\label{fig:comp}
\end{figure}

Figure \ref{fig:comp} compares the lower limits derived here with current observational upper limits, regions of theoretical interest, and sensitivities of planned experiments.
Only the lower limits for the \FRV~are shown, since they all lie below the limits derived with the \KD.
The lower limits clearly extend below the stringent upper limits from the CAST experiment \cite{cast2007} (dark shaded region).
In the \scenIGMF~case, they also lie below the upper limit derived from the nonobservation of prompt \gray s from the supernova SN\,1987a (gray shaded region) \cite[][]{brockway1996,grifols1996}. 
These \gray s would be the result of ALPs reconverted in the GMF that are produced in the supernova explosion
\footnote{The limits should be considered as an order of magnitude estimate since they rely on some simplifications.
For instance, a constant \pa~conversion probability in the GMF is assumed.}.
The dotted-dashed lines show theoretical upper limits on $\gag$ calculated from magnetic white dwarfs (mWDs) \cite[][]{gill2011}.
\Pa~conversions lead to a linear polarization $P_L$ of the photon beam \cite{raffelt1988}, and by treating the current observations of mWDs as a limit, i.e., 
$P_L \lesssim 5\,\%$,  one can derive a limit on the \pa-coupling.
The different lines correspond to different values of the magnetic field strength of the mWDs and different values for the limit on $P_L$.
Although the magnetic field and ambient density in the vicinity of mWDs are very different from the scenarios considered here,
the mWD considerations turn out to be sensitive in the same $(\ma,\gag)$ region as the VHE observations.
Nevertheless, the limits use a $B$-field model inferred from one single mWD. 
If they are confirmed with future observations, they will strongly constrain the parameter space for ALPs that can potentially decrease the opacity of the Universe for VHE \gray s.

The lower limits of the optimistic scenarios extend into the preferred region for the ALP parameters
to explain the white dwarf (WD) cooling problem.
It is difficult with current theoretical models to satisfactorily reproduce the observed WD luminosity function.
Including the production of ALPs, on the other hand, with a mass and coupling within the light-gray-shaded band in Fig. \ref{fig:comp}
\footnote{
Reference \cite{isern2008} set bounds on the mass of the QCD axion or equivalently on the electron-axion coupling.
This can be translated into a bound on the photon-axion coupling \cite{raffelt2008,redondo2013} and consequently on the
photon-ALP coupling. The values shown here are taken from \cite{hewett2012}
}
serves as an additional cooling mechanism for WD and 
can reduce the tension between current model predictions and data \cite{isern2008}.
This issue is, however, subject to ongoing discussion \cite{melendez2012}.

Interestingly, the ALP parameter space of the \scenfid-scenario can be probed with planned experiments. 
The sensitivity forecasts for the improved Any Light Particle Search (ALPS II) \cite{hewett2012,alpsII}
and the International Axion Observatory (IAXO) \cite{irastorza2011} are displayed as a crosshatched and righthatched region, respectively, in Fig. \ref{fig:comp}.
The lower limits derived here thus pose an additional physics case for these future experiments.

\section{Summary and Conclusions}
\label{sec:conclusion}

In this article, VHE-\gray~observations have been used for the first time, to set lower limits on the \pa~coupling.
Various magnetic field configurations have been considered, namely the magnetic field of the Milky Way, the intergalactic magnetic field,
and the $B$ field that pervades the intra-cluster medium.
Under the assumption of domain-like random $B$ fields, the \pa-conversion probability for different field strengths and coherence lengths is investigated.
As a result, hypothetical ALPs from a large region of the $(\ma, \gag)$ parameter space would be able to reduce the opacity of the Universe for VHE-\gray s and to ease the tension 
between the predictions of common EBL models and data from IACTs that has been found in Ref. \cite{horns2012}.
If field strengths and the coherence lengths of the magnetic fields are assumed to be close to their maximally observationally allowed values,
lower limits on $\gag$ are obtained that reach down to $\gag \sim 10^{-12}\mathrm{GeV}^{-1}$. 
Even lower values of $\gag$ could be obtained if, for instance, the GMF model were tuned to the most optimistic values allowed by the fitting errors calculated in Ref. \cite{jansson2012}.
Additional contributions to the GMF are also possible, such as a kiloparsec-scale magnetized wind \cite{everett2008} that could further enhance the 
conversion probability and reduce the lower limits.
For more conservative values of the intervening $B$ fields, the limits are of the order of $\gag \gtrsim 2\times10^{-11}\mathrm{GeV}^{-1}$,
close to the upper bounds of the CAST experiment and within the sensitivity estimates of future experiments such as ALPS II or IAXO.

Alternative mechanisms that are, in principle, able to increase the transparency of the Universe are also discussed in the literature.
Active galactic nuclei could also be the source of cosmic-ray protons that produce secondary photons in the interaction with the EBL \cite{essey2010}.
In such scenarios, the secondary photons are responsible for the high-energy end of VHE \gray~spectra. 

The effect of the unknown EBL density on the lower limits of $\gag$ has been assessed with two different EBL models.
At the time being, the sample of VHE spectra is dominated by sources with a redshift $0.1\lesssim z \lesssim 0.2$ and is thus most sensitive to changes in the EBL density at near-infrared wavelengths.
As a consequence, certain EBL model realizations exist for which the pair production anomaly is less significant \cite{meyer2012ppa},
and higher values of $\gag$ would be required to obtain a significant improvement over the situation without ALPs.
One improvement would be to parametrize the EBL model independently 
(for instance, with splines, as done in, e.g., Ref. \cite{meyer2012}) and recalculate the significances in the presence of ALPs. 
This is left for future investigations.
Firm conclusions can only be drawn with future direct observations of the EBL and VHE measurements
in the optical thick regime of both distant sources at several hundreds of GeV and nearby sources at several tens of TeV, 
which will also enable further tests of ALP scenarios. 
Several such observations have already been announced
\footnote{For example, the detection of the distant BL Lac KUV\,00311-1938 with H.E.S.S. or the observation of a flaring state of Markarian\,421 with VERITAS were recently announced, see {http://tevcat.uchicago.edu/}}
and will become more feasible with the next generation of air shower experiments such as the Cherenkov Telescope Array \cite{cta2011}, the High Altitude Water Cherenkov Experiment \cite{hawc2005}, 
and the Hundred*i Square-km Cosmic ORigin Explorer \cite{hiscore2011}. 

\acknowledgements
M.M. would like to thank the state excellence cluster ``Connecting Particles with the Cosmos'' at the University of Hamburg.
The authors would also like to thank Alessandro Mirizzi and the anonymous referees for helpful comments improving the manuscript.

\begin{appendix}

\section{Fit qualities}
\label{app:fit_qual}
In this appendix, the quality of the fits of the analytical functions introduced in Eq. \ref{eq:fitfunc} is addressed. 
The fit statistics without the contribution of ALPs are listed in Table \ref{tab:fit_qual}. 
For both EBL models, all spectra show a high fit probability. The only exceptions are the spectra of Mrk\,501 and one spectrum of 3C\,279.
The former spectrum is measured with high accuracy resulting in very small statistical errors. 
It is dominated by its systematical uncertainties (see gray band in Fig. 10 in \cite{mkn501hegra1999}) which are not included in the fit here. 
The spectrum of 3C\,279 consist only of three data points, making a power law the only meaningful fitting function.
The combined $\chi^2$ values translate into satisfactory overall fit probabilities of $p_\mathrm{fit} = 0.160$ and $p_\mathrm{fit} =  0.303$ for the \KD~and 
\FRV, respectively.

\begin{table}[t]
 \centering
 \begin{tabular}{llcc|cc|cc}
  \hline
  \hline
 \multirow{2}{*}{$j$} & \multirow{2}{*}{Source} & \multirow{2}{*}{Experiment} &Fit  & $\chi^2$ (d.o.f.) & $p_\mathrm{fit}$ & $\chi^2$ (d.o.f.) & $p_\mathrm{fit}$ \\
     {} & {} & {}& function\footnotemark[1] & \multicolumn{2}{c|}{$\tau = 1 \times\,\tau_\mathrm{KD}$} & \multicolumn{2}{c}{$\tau = 1.3 \times\,\tau_\mathrm{FRV}$} \\ 
  \hline
  1 & Mrk\,421 & HEGRA & LP & $\ldots$ & $\ldots$ & 4.10 (7) & 0.768 \\
  2 & Mrk\,421 & HEGRA & PL & $\ldots$ & $\ldots$ & 8.75 (8) & 0.364 \\
  3 & Mrk\,421 &  \hess & LP & $\ldots$ & $\ldots$ & 14.75 (10) & 0.142 \\
  4 & Mrk\,421 &  \hess & LP, PL\footnotemark[2] & 16.97 (11) & 0.109& 13.95 (12) & 0.304 \\
  5 & Mrk\,501 &  HEGRA & LP&  29.78 (14) & 0.008 & 38.55 (14)  & 0.000 \\
  6 & 1ES\,1950+650  & HEGRA & PL & $\ldots$ & $\ldots$ & 0.78 (3) & 0.854 \\
  7 & 1ES\,1950+650  & HEGRA & PL & $\ldots$ & $\ldots$ & 7.89 (6) & 0.247 \\
  8 & PKS\,2155-304\ & \hess & PL& $\ldots$ & $\ldots$ & 9.57 (7) & 0.215  \\
  9 & PKS\,2155-304\ & \hess & PL& $\ldots$ & $\ldots$ & 8.34 (8) & 0.401  \\
  10 & PKS\,2155-304 & \hess & LP& $\ldots$ & $\ldots$ & 11.48 (14) & 0.648 \\
  11 & PKS\,2155-304 & \hess & PL& $\ldots$ & $\ldots$ & 5.46 (3) & 0.141  \\
  12 & RGB\,J0710+591& VERITAS&PL& $\ldots$ & $\ldots$ & 2.00 (3) & 0.573 \\
  \multirow{2}{*}{13} & \multirow{2}{*}{H\,1426+428}& HEGRA, & \multirow{2}{*}{PL} & \multirow{2}{*}{9.06 (10)} &\multirow{2}{*}{0.526} & \multirow{2}{*}{9.33 (10)} &\multirow{2}{*}{0.501} \\
  {} & {} &CAT, WHIPPLE & {} & {} & {} & {} & {}  \\
  14 & 1ES\,0229-200& \hess & PL & 3.17 (6) & 0.787 & 3.86 (6) & 0.695 \\
  15 & H\,2356-309 & \hess & PL & $\ldots$  &$\ldots$   & 3.56 (6) & 0.735 \\
  16 & H\,2356-309 & \hess & PL & $\ldots$  & $\ldots$  & 4.20 (6) & 0.649 \\
  17 & H\,2356-309 & \hess & PL & $\ldots$  & $\ldots$  & 3.70 (6) & 0.717 \\
  18 & 1ES\,1218+304 & VERITAS & PL & $\ldots$ & $\ldots$ & 2.33 (5) & 0.802 \\
  19 & 1ES\,1101-232 & \hess & PL & 5.70 (11) & 0.892 & 6.83 (11) & 0.813 \\
  20 & 1ES\,0347-121 & \hess & PL & 2.59 (5) & 0.763  & 2.30 (5) & 0.807  \\
  21 & RBS\,0413 & VERITAS& PL & $\ldots$& $\ldots$ & 0.02 (2) & 0.988 \\
  22 & 1ES\,0414+009 & \hess & PL & 1.86 (4) & 0.761  & 2.95 (4) & 0.567 \\
  23 & 1ES\,0414+009 & VERITAS & PL & $\ldots$ & $\ldots$ & 0.51 (2) & 0.776 \\
  24 & PKS\,1222+21  & MAGIC & PL & $\ldots$ & $\ldots$ & 0.24 (3) & 0.971 \\
  25 & 3C\,279 & MAGIC & PL & 3.46 (1) & 0.063 & 3.95 (1) & 0.047 \\
  26 & 3C\,279 & MAGIC & PL & 3.64 (3) & 0.303 & 4.48 (3) & 0.214 \\
\hline
\multicolumn{4}{l|}{\textbf{Combined}} & 76.25 (65)  & 0.160 & 173.88 (165)& 0.303 \\
\hline
 \end{tabular}
\caption{List of fit qualities for all VHE-\gray~spectra if no ALPs are included in the de-absorption of the spectra. The table shows the $\chi^2$ values,
 the degrees of freedom (d.o.f.), and the resulting fit probabilities $p_\mathrm{fit}$ for both EBL models used here.
See Table \ref{tab:spectra} for the references of each spectrum.
}
\label{tab:fit_qual}
\footnotetext[1]{PL = power law, LP = logarithmic parabola.}
\footnotetext[2]{The spectrum is fitted with a logarithmic parabola in the \KD~and with a power law in the \FRV, respectively.}
\end{table}

The fit qualities for the de-absorption of the observed VHE spectra with ALPs 
are poor in the transition to the strong mixing regime with $p_\mathrm{fit} \ll 1$.
The corresponding high $\chi^2$ values are dominated by a few spectra only (as in the no-ALPs case), namely those with high count statistics and consequently small error bars. 
The largest contributions come from the spectra of Mrk\,421 \cite{mkn421hess2010} and Mrk\,501 \cite{mkn501hegra1999}, for the \KD, and, 
additionally, the spectrum of PKS\,2155-304 \cite{pks2155hess2007} for the \FRV. 
Again, these spectra are dominated by their systematic uncertainties which are not included here.

The reason for the large contribution of these spectra to the total $\chi^2$ values is the oscillatory 
behavior outside the SMR.
As a result of the oscillations, the fit residuals also scatter strongly around zero and give rise to 
a low fit probability. 
If the spectra with high statistics are removed from the samples, the $\chi^2/\mathrm{d.o.f.}$ values are close to one for all $(\ma,\gag)$ values.

To summarize, the fits to most spectra in the $(\ma,\gag)$ parameter space are acceptable. 
In the case of a small overall fit probability, the lower limits on $\gag$ are pushed towards lower
values as the residual distribution broadens. 
Nevertheless, this bias mainly affects the region of ALP parameters outside the SMR and could be eliminated if the systematic uncertainties were included in the fit.

\end{appendix}

\bibliography{meyer_ALP_transparency,vhe_spectra}
\end{document}